\documentclass[prc,preprint,nofootinbib,preprintnumbers,longbibliography,superscriptaddress,tightenlines,usletter]{revtex4-2}

\usepackage{amsmath,mathrsfs}    
\usepackage{graphicx}   
\usepackage[caption=false]{subfig}
\usepackage{color}
\usepackage[colorlinks=true,linkcolor=blue,citecolor=blue, urlcolor=blue]{hyperref}   
\usepackage{bm} 
\usepackage{ulem}
\usepackage[inline]{enumitem}
\usepackage{soul}
\usepackage{graphicx}
\usepackage{amsmath,amssymb}
\usepackage{soul}
\usepackage{textcomp}
\usepackage{hyperref}
\usepackage[toc,page]{appendix}
\usepackage{xcolor}
\usepackage{tikz}

\usepackage{array}   
\newcolumntype{C}{>{$}c<{$}} 

\definecolor{MyDarkBlue}{RGB}{22, 79, 134}
\definecolor{MyLightRed}{RGB}{200, 37, 6}

\newcommand{\QD}{\mathfrak{Q}_D}

\def\x{{\bm x}}

\def\k{{\bm k}}

\def\k{{\bm k}}
\def\F{{\mathcal H}}

\def\st{\begin{equation}}
\def\stp{\end{equation}}
\def\bg{\begin{eqnarray}}
\def\nd{\end{eqnarray}}
\def\Eq#1{eq.~(\ref{#1})}

\def\eq#1{(\ref{#1})}
\def\app#1{App.~\ref{#1}}
\def\Fig#1{Fig.~\ref{#1}}

\def\Sect#1{Sect.~\ref{#1}}

\def\llangle{\left\langle}
\def\rrangle{\right\rangle}
\def \bes {\begin{subequations}}
\def \ees {\end{subequations}}

\def \F
\def\chemconst{{\chi_0}}

\newcommand{\App}[1]{{App.~\ref{#1}}}

\newcommand{\dd}{\mathrm{d}}

\begin{document}

\title{Dynamics of the $O(4)$ critical point in QCD: critical pions and diffusion in Model G}
\author{Adrien Florio}
\email[]{aflorio@bnl.gov}
\affiliation{Department of Physics, Brookhaven National Laboratory, Upton, New York 11973-5000, USA}
\author{Eduardo Grossi}
\email[]{eduardo.grossi@unifi.it}
\affiliation{Dipartimento di Fisica, Universit\`a di Firenze and INFN Sezione di Firenze, via G. Sansone 1,
50019 Sesto Fiorentino, Italy}
\author{Derek Teaney}
\email[]{derek.teaney@stonybrook.edu}
\affiliation{Center for Nuclear Theory, Department of Physics and Astronomy, Stony Brook University, New York 11794-3800, USA}

\date{\today}
   \begin{abstract}
We present a detailed study of the finite momentum dynamics of the $O(4)$
critical point of QCD, which lies in the dynamic universality class of ``Model
G".  The critical scaling of the model is analyzed in multiple dynamical
channels. For instance, the finite momentum analysis allows us to precisely
extract the pion dispersion curve below the critical point.  The pion velocity
is in striking agreement with the predictions relation and static universality.
The pion damping rate and velocity are both consistent with the dynamical
critical exponent $\zeta = 3/2$ of ``Model G".  Similarly, although the
critical amplitude for the  diffusion coefficient of the conserved $O(4)$
charges is small, it is clearly visible both in the restored phase and with
finite explicit symmetry breaking,  and its dynamical scaling is again
consistent with  $\zeta=3/2$.  We determine a new set of universal dynamical
critical amplitude ratios relating the diffusion coefficient to a suitably
defined order parameter relaxation time. We also show that in a finite volume
simulation, the chiral condensate diffuses on the coset manifold in a manner
consistent with dynamical scaling, and with a diffusion coefficient that is
determined by the transport coefficients of hydrodynamic pions.  Finally, the
amplitude ratios (together with other non-universal amplitudes also reported
here) compile all relevant information for further studies of ``Model G" both
in and out of equilibrium.

   \end{abstract}



\maketitle

\clearpage


\section{Introduction}
\label{sec:intro}

High-multiplicity reactions in colliders are a unique tool to explore the phase diagram of QCD. Their ability to probe different phases and phase transitions crucially relies on a precise understanding of its dynamics in a hot and dense state. Arguably one of the most impactful results of the scientific program of the Relativistic Heavy-Ion Collider (RHIC) was the discovery of a very short thermalization time after which  the quark-gluon plasma created in the collision exhibits collective behavior, faithfully reproduced by hydrodynamics. This phenomenon is key to the predictive power of heavy-ion collisions.

This work is a direct continuation of \cite{Florio:2021jlx} and touches on the nature of the appropriate hydrodynamic theory required by heavy-ion collisions. Away from a phase transition, the hydrodynamics of a system is fully characterized by symmetries and conserved charges. At criticality, the dynamics of the order parameter and its coupling to the conserved charges also needs to be considered~\cite{Hohenberg:1977ym}.
This is crucial to the dynamics of two-flavor QCD in the chiral limit, close to
its second order phase transition~\cite{Rajagopal:1992qz}. In  the chiral
limit, the hydrodynamic theory above $T_c$ is specified by an unbroken
$SU_L(2) \times SU_R(2)\simeq O(4)$ symmetry group with corresponding conserved
charges. Below $T_c$ the symmetry is broken to $SU_V(2) \simeq O(3)$ and the
associated massless Goldstone modes (the pions) must be incorporated into the hydrodynamic
effective theory~\cite{Son:1999pa,Son:2002ci,Son:2001ff}.
Finally, in the critical region the appropriate macroscopic description includes an $O(4)$ order parameter ($\phi_a =
(\sigma, \vec{\pi})$), and the effective theory interpolates between the two hydrodynamic limits above and below the critical
point.


Beyond the theoretical appeal of the chiral limit, the real-world up and down quark masses are small compared to QCD scales, to the point where it is legitimate to ask whether aspects of real-world QCD close to its phase transition can be recovered by deforming the massless limit. For static quantities this question was asked in \cite{Pisarski:1983ms,Wilczek:1992sf} and led to further studies of the chiral critical point with explicit symmetry breaking~\cite{Toussaint:1996qr,Engels:2009tv,Engels:2011km,Engels:2014bra,Butti:2003nu,Braun:2009ruy,Braun:2010vd,Braun:2011iz,Braun:2020ada}. These studies, when combined with important results from Lattice QCD at various quark masses~\cite{HotQCD:2019xnw,Kotov:2021rah,Cuteri:2021ikv,Kaczmarek:2020sif}, indicate that the static properties of the chiral crossover in real world QCD can be reasonably understood using universality considerations and the existence of a critical point in the massless two-flavor limit.

It remains to be seen whether the dynamics of the chiral crossover can be similarly understood using the proximity of QCD to the $O(4)$ critical point.
The relevant hydrodynamics of the  phase transition lies in the dynamical universality class of  ``Model G" in the Halperin-Hohenberg classification scheme~\cite{Rajagopal:1992qz},  which  predicts the existence of long-range chiral modes that describe the evolution of soft pions in the crossover region.
When the $O(4)$ symmetry is explicitly broken, the pions develop a mass, tempering the impact of these modes on the  dynamics on the system.
Nevertheless, although these modes contribute negligibly to the energy momentum tensor, they survive over time scales much longer than the typical relaxation time and most definitely affect the long-range dynamics of chiral observables  in QCD in the crossover region.

For an experimental perspective, although  it is challenging to systematically measure soft pions of transverse momenta  smaller than  $400\,{\rm MeV}$, recent analyses uncovered some discrepancies between state-of-the-art hydrodynamic models and  available data in this momentum range. The results are indicative that conventional hydrodynamics does not fully describe the long wavelength dynamics of these pseudo-Goldstone modes  \cite{Nijs:2020ors, Mazeliauskas:2019ifr, Devetak:2019lsk}.

Taking these indications into consideration, and further encouraged by the proposed upgrade to the ALICE detector \cite{ALICE} which will be more
sensitive to low $p_T$ pions, a natural question, already raised in \cite{Rajagopal:1992qz,Son:2001ff} and subsequently studied further~\cite{Grossi:2020ezz,Grossi:2021gqi}, is the following: what is  the effect of the critical chiral dynamics on the hydrodynamics of real-world QCD? To answer this question a more precise and detailed understanding of the dynamics of ``Model G" itself is required.

Over the years, a series of classical-statistical simulations have been presented in the literature \cite{Berges:2009jz,Tauber:2016mpa,Schlichting:2019tbr,Schweitzer:2020noq,Schweitzer:2021iqk,Schaefer:2022bfm,Chattopadhyay:2023jfm}, building up towards  studies of the dynamics of ``Model G" and ``Model H", the critical model of the conjectured second order phase transition at the finite baryon chemical potential~\cite{Son:2004iv}. In a similar spirit, functional Renormalization Group (fRG) studies have been carried out in \cite{Canet:2003qd,Benitez:2009xg, DePolsi:2021cmi,Roth:2023wbp,Floerchinger:2011sc,Floerchinger:2016gtl,Dupuis:2020fhh}.  Although limitted to mean field, holographic models can provide additional analytical insight into the dynamics of the chiral phase transition~\cite{Cao:2022csq}. 

 This work expands on \cite{Florio:2021jlx} (where the first successful
 numerical simulations of ``Model G" were presented) by extending the zero
 momentum analysis  to finite momenta. With this extension, we are able to
 quantify the dispersion curve of critical pions in detail and study the
 critical diffusion of the conserved charges. We show that in a finite volume
 simulation the orientation of the chiral condensate diffuses on the three sphere (the coset manifold), with a diffusion coefficient determined by the transport coefficients of hydrodynamic pions in infinite volume.
 When 
 chiral symmetry is explicitly broken, the diffusive dynamics is coupled to
 a Hamiltonian structure determined by the pion mass, and the combined evolution dynamically realizes the $\epsilon$-regime of QCD at finite temperature\footnote{ 
The $\epsilon$ regime is when the quark mass $m_q\propto H$ is small and the volume $V$ is large, but $H\bar \sigma(T) V/T\sim 1$. Here $\bar \sigma(T)$ is the chiral condensate as a function of temperature -- see \cite{Damgaard:2011gc} for an overview. }.
 We also compute new universal dynamical ratios relating the diffusion coefficient to the
 (appropriately defined) order parameter relaxation time, which we also
 determine both above $T_c$ and on the critical line. In total, these
 measurements corroborate the dynamical scaling of ``Model G" with critical
 exponent $\zeta = d/2$  across multiple observables, ranging from the pion
 velocity and damping rate, to the diffusion of the $O(4)$
 charges\footnote{Here $d=3$ is the dimensionality of space.}. These results
 will prove essential to the next step of this series of work, which will
 quantitatively analyze the non-equilibrium dynamics of the expanding critical
 system.

\section{Overview of Model G and its phases}

\subsection{Model G}

The theory we are considering is a generalization of ``Model G" of \cite{Hohenberg:1977ym} and was first introduced in \cite{Rajagopal:1992qz}. It is the relevant hydrodynamic theory constructed around the $O(4)$ critical point of ``2-flavor chiral QCD", namely QCD with massless up and down quarks ($m_u=m_d=0$). It contains a $O(4)$ order parameter $\phi_a\equiv(\sigma, \vec \pi)$, a proxy for the quark condensate of real QCD, dynamically coupled to vector  and axial charges\footnote{
Here $a$ and $b$ denote $O(4)$ indices; $s$, $s_1$, $s_2$, etc. denote the three isospin indices, i.e. the components of $\vec{\pi}=(\pi_1, \pi_2, \pi_3)$;
finally, spatial indices are notated $i$, $j$ and $k$. The dot product indicates an appropriate contraction of
indices when clear from context, e.g. $\phi\cdot \phi=\phi_a \phi_a$,
$\vec{\pi} \cdot \vec{\pi} = \pi_s \pi_s$, and $\nabla \cdot \nabla =\partial_i
\partial^i$. }, $n_V^s$  and $n_{A}^s$ respectively. They correspond to the
original iso-vector and iso-axial vector currents, $\vec n_{V}\sim \bar q
\gamma^0 \vec t_I q$ and  $\vec n_{A}\sim \bar q \gamma^0\gamma^5 \vec t_I q$,
respectively. These can be conveniently combined into an antisymmetric tensor
of charge densities $n_{ab}\in\mathfrak{o}(4)$:
\begin{align}
(n_V)_{s} &= \frac12 \epsilon_{0s s_1 s_2} n_{s_1s_2} \, ,  \\
(n_A)_s   &= n_{0s} \, .
\end{align}
The free energy (or effective Hamiltonian)
in the presence of an explicit symmetry breaking $H_a=(H,\vec 0)$  can be studied using an $O(4)$ Landau-Ginzburg action
\st
\label{eq:Hdef}
\mathcal H \equiv  \int d^3x \left[ \frac{n^2}{4 \chi_0}    +\frac{1}{2} \partial_i \phi_a \partial^i \phi_a + V(\phi) - H \cdot \phi  \right]\, ,
\stp
where
\st
V(\phi) = \frac{1}{2} m_0^2 \, \phi^2  + \frac{\lambda}{4} (\phi\cdot\phi)^2 \, ,
\stp
with $m_0^2$ negative. In equilibrium, the charges  have Gaussian fluctuations set by a susceptibility $\chi_0$, which is the same coefficient for both the iso-vector and the iso-axial-vector charges near the critical point. The interactions between the charges and  the order parameter enters the dynamics through the equations of motion and are determined by the non-trivial Poisson brackets between these fields -- see \cite{Rajagopal:1992qz,Grossi:2021gqi} for explicit expressions and a derivation. This construction leads  to the following evolution equations 
\begin{subequations}
\begin{align}
   \partial_t \phi_a  + \frac{g}{\chi_0}\,n_{ab} \phi_b  &= -\Gamma_0 \frac{\delta \mathcal H}{\delta \phi_a} + \theta_a \, , \label{eq:eom1}\\
   &=\Gamma_0 \nabla^2 \phi_a - \Gamma_0 (m_0^2 + \lambda \phi^2)\phi_a + \Gamma_0 H_a + \theta_a\, ,\\
   \partial_t n_{ab}  + g \,\nabla \cdot (\nabla \phi_{[a} \phi_{b]}) + H_{[a}  \phi_{b]} &= \sigma_0 \nabla^2\frac{ \delta \mathcal H }{\delta n_{ab} }    +  \partial_{i} \Xi_{ab}^i  \, , \label{eq:eom2}\\
   &= D_0 \nabla^2 n_{ab} +  \partial_{i} \Xi_{ab}^i \label{eq:eom3} \, .
\end{align}
\end{subequations}
Here, for example,   $H_{[a}\phi_{b]}$ denotes the anti-symmetrization, $H_a \phi_b - H_b \phi_a$. The coefficients  $\Gamma_0$ and $\sigma_0$  are the bare kinetic coefficients associated with the order parameter and the charges. The bare diffusion coefficient of the charges is $D_0 = \sigma_0/\chi_0$. The constant $g$ is a coupling of the field $\phi$, and has the units of $({\rm action})^{-1}$ in our conventions.
Finally, $\theta_a$ and $\Xi_{ab}$ are the appropriate  noises, which  are defined through their two-point correlations~\cite{Hohenberg:1977ym}
\begin{subequations}
\begin{align}
   \langle \theta_a(t,x)\theta_b(t',x') \rangle &= 2 T_c\Gamma_0 \, \delta_{ab} \, \delta(t-t')\delta^3(x-x') \, ,\label{eq:langevin_var}\\
   \langle \Xi^i_{ab}(t,x)\Xi^j_{cd}(t',x') \rangle &= 2  T_c\sigma_0 \, \delta^{ij} \left(\delta_{ac} \delta_{bd} - \delta_{ad} \delta_{bc} \right) \, \delta(t-t')\delta^3(x-x') \ . \label{eq:langevin_var_cons}
\end{align}
\end{subequations}
The dissipative Model G dynamics  relaxes to the equilibrium distribution for the fields $\phi_a$ and $n_{ab}$~\cite{Hohenberg:1977ym}
\st
  P[\phi, n] = \frac{1}{Z} e^{-\mathcal H[\phi, n]/T_c } \, ,
\stp
which reproduces the static critical behavior of the $O(4)$ model.

The basic observable that will be analyzed  are statistical two-point function between
fields as a function of relative momentum and time.
In the presence of the explicit breaking $H_a=H \delta_{a0}$ the flavor index can always be
decomposed in a longitudinal $\sigma$ and transverse $\vec{\pi}$ direction, therefore it is natural to define:
\begin{align}
G_{\sigma\sigma}(t,\k)= \frac{1}{V} \langle \phi_0(t,\k)\phi_0(0,-\k) \rangle_c \, ,
\end{align}
for the scalar channel,  and
\begin{align}
G_{\pi\pi}(t,\k)= \frac{1}{3V} \sum_{s=1}^3\langle \pi_s(t,\k)\pi_s(0,-\k) \rangle_c \, ,
\end{align}
for the pseudo-scalar channel.
The statistical correlators for the charges are defined analogously
\begin{align}
G_{AA}(t,\k)&= \frac{1}{3V} \sum_{s=1}^3\langle n_{As}(t,\k) n_{As}(0,-\k) \rangle_c\, , \\
G_{VV}(t,\k)&=\frac{1}{3V} \sum_{s=1}^3\langle n_{Vs}(t,\k) n_{Vs}(0,-\k) \rangle_c \ .
\end{align}
Note that all these correlators can be studied in mean field, see
\cite{Grossi:2021gqi}.

At zero explicit breaking  $H=0$, we will also consider  isotropized $O(4)$ charge and field correlators
\begin{align}
G_{\phi\phi} (t,\k) &= \frac{1}{4V}\sum_{a}\langle \phi_a(t,\k) \phi_a(0,-\k) \rangle_c,\label{eq:broken_cor1}
\\
 G_{nn} (t,\k) &=\frac{1}{6V}\sum_{a>b} \langle n_{ab}(t,\k) n_{ab}(0,-\k) \rangle_c \ .
\label{eq:broken_cor2}
\end{align}

\subsection{Simulations}

This paper is a continuation of \cite{Florio:2021jlx} and presents results obtained by numerically solving equations \eqref{eq:eom1}-\eqref{eq:eom3}, using an algorithm detailed in Appendix A of \cite{Florio:2021jlx}. The stepper combines a symplectic integrator for the Hamiltonian evolution and Metropolis updates for the dissipative step.
Throughout we will present our results in lattice units, setting
$g=T_c=1$, and setting the lattice spacing to unity, $a=1$. This amounts to setting $\hbar=1$ and $T_c=1$ and choosing the microscopic lattice spacing so that the model reproduces the correlation length of the physical system. The pole frequency of the pion near the critical point is reproduced by this choice of units. The quartic coupling is set to $\lambda=4$  and the model is simulated close to its critical mass, $m_c^2(\lambda) = -4.8110(4)$.
The reduced temperature and magnetic field are defined as in \cite{Engels:2014bra} :
\st
          t_r \equiv \frac{m_0^2 - m_c^2}{|m_c^2|} \, ,   \qquad   h \equiv \frac{H}{H_0} \, ,
\stp
where $H_0=5.15(5)$ is chosen so that  the magnetization scales as
\st
        \bar \sigma   = h^{1/\delta}  \, ,
\stp
on the critical line, $t_r=0$.
For $H=0$ and in the broken phase  the magnetization scales as
\st
        \bar \sigma   = B^{-} (-t_r)^{\beta} \ ,
\stp
with amplitude\footnote{Note an unfortunate misprint in \cite{Florio:2021jlx}. Equation (29) in the
published version should read
$\Sigma=b_1(m_c^2-m^2)^\beta(1+C_T(m_c^2-m^2)^{\omega\nu})$. The then quoted
parameters $b_1=0.544\pm0.004$ and $C_T =0.20\pm0.02$ lead to a parametrization
        $\bar \sigma   = B^{-} (-t_r)^{\beta}(1 + B_1 (-t_r)^{\nu\omega} )$ ,
with $B^-=|m_c^2|^\beta b_1 = 0.988\pm0.007$ and $B_1=C_T|m_c^2|^{\omega\nu} = 0.49\pm0.05$. },
$B^{-}=0.988(7)$. The non-universal parameters $B^{-}$ and $H_0$ completely fix the properties of the magnetic equation of state close to the critical point.

For the static critical exponents quoted here and below we take the values $\beta=0.380(2)$ and $\delta=4.824(9)$ from ~\cite{Engels:2014bra}.
For the correction-to-scaling exponent we take $\omega=0.755(5)$~\cite{Hasenbusch:2021rse} and the for the dynamical critical exponent
we take $\zeta=d/2$~\cite{Rajagopal:1992qz},  where $d=3$ is the number of spatial dimensions. A summary of all exponents used in this work is given in Table~\ref{tab:num_values_fixed} of \App{Sec:StaticMeasurements}.

We performed additional analyses  along the critical line using the data generated in \cite{Florio:2021jlx} for different lattice sizes $L$.
We also generated an extensive set of data at $H=0$, both in the broken ($m^2_0 < m_c^2$) and restored ($m^2_0> m_c^2$) phases.
The dynamical parameters were chosen as in our previous work, $\chi_0 = 5$, $\Gamma_0=1$ and $D_0=1/3$. Our full dataset is summarized in Table~\ref{tab:sims}.

\begin{figure}
\includegraphics{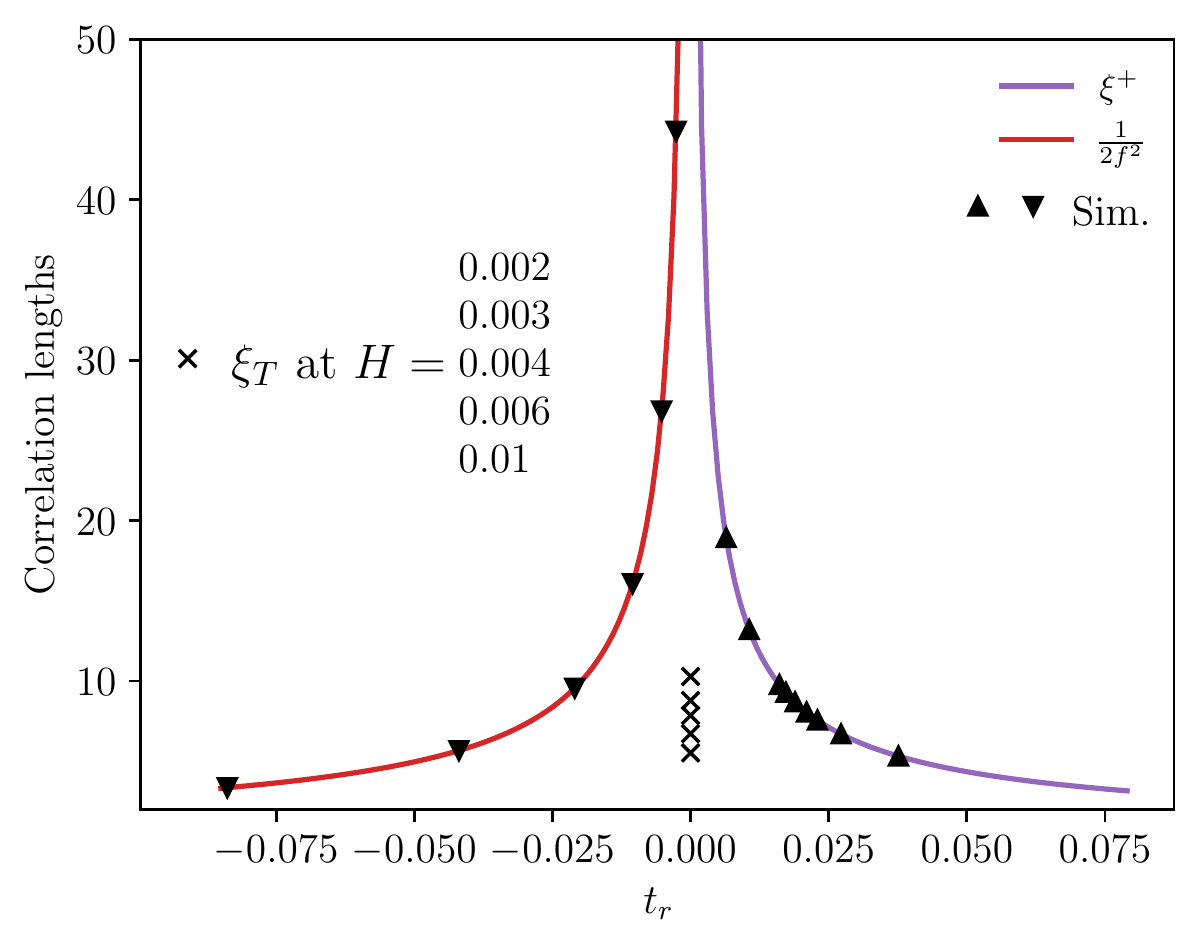}
\caption{The correlation length versus temperature in our simulations close to the $O(4)$ critical point. The correlation length in the restored phase and on the critical line is obtained from  fits described in App.~\ref{app:corrlength_unbroken} and~\ref{app:corrlengthcrit}. We use $T/2f^2$ as a proxy for the correlation length in the broken phase, with $T=1$ in our adopted units (see text).
}
\label{fig:recap}
\end{figure}

For context, we also display the correlation length (in lattice units) for our different simulations in Fig.~\ref{fig:recap}.
The correlation lengths in the restored phase and on the critical line correspond to the fits presented in App.~\ref{app:corrlength_unbroken} and~\ref{app:corrlengthcrit}. As discussed below, in the broken phase the hydrodynamic theory is characterized by a pion EFT at the longest wavelengths.
The pion parameter $f^2(T)$ is a matching coefficient in the EFT, which  characterizes the pion's fluctuations and scales as the inverse correlation length
close to the critical point~\cite{Hasenfratz:1989pk,Son:2002ci}.  We will use
$1/{(2f^2)}$ as an estimate for the correlation length in the broken phase
which is determined numerically in \App{app:staticanalysis}. Here  the factor of $\tfrac{1}{2}$
is conventional and is motivated by the fact that finite volume corrections to
correlation functions (which are computable using the pion
EFT~\cite{Hasenfratz:1989pk}) are organized in powers of $1/(2f^2L)$ with order
one coefficients, see App.~\ref{app:staticanalysis}.
The black data points correspond to the different temperatures we simulated.
\begin{table}
  \centering
  \begin{tabular}{|c|c|c|c|c|}
    \hline
    Phase&$m^2_0$&$t_r$&$L$&$H$\\
    \hline
    Restored &\begin{tabular}{ccccc}
      -4.6300 & &-4.7100 &  & -4.7337\\
      -4.6800 & &-4.7200 &  & -4.7600\\
      -4.7005 & &-4.7280 &  & -4.7800
    \end{tabular}&
    \begin{tabular}{ccccc}
     0.03762& & 0.02099&  & 0.01608\\
     0.02723 & & 0.01891 &  & 0.01060 \\
     0.02296 & & 0.01725 &  & 0.006444
   \end{tabular} & $96$ & $0$ \\
    \hline
    Critical line & \centering $m^2_c=-4.8110$ & 0 & 80 &  \begin{tabular}{ccc}
    0.002 & 0.006 \\
    0.003 & 0.01 \\
    0.004 &
    \end{tabular}\\
    \hline
    Broken & \begin{tabular}{ccccc}
    -4.8236 & &-4.9118 & \\
    -4.8362 & &-5.0127 &\\
    -4.8614 & &-5.2143 &
  \end{tabular}& \begin{tabular}{ccccc}
  -0.002620 & & -0.02096 & \\
  -0.005239 & & -0.04191 &\\
  -0.01048 & & -0.08383 &
\end{tabular} &  \begin{tabular}{ccc}
    48 & 128 \\
    64 &  \\
    96 &
    \end{tabular}  & 0 \\
    \hline
  \end{tabular}
  \caption{Dataset analyzed in the present work. For all simulations, we fix $\lambda=4$, $\chi_0 = 5$, $\Gamma_0=1$ and $D_0=1/3$. $L$ is the size of our lattices.}
  \label{tab:sims}
\end{table}

Before diving into a quantitative study, we use the rest of this section to present the qualitative behavior of the different channels in the restored and broken phases (with $H=0$)   and on the critical line ($t_r=0$ and $H\neq0$).

\subsection{The restored phase with $H=0$}

\begin{figure}
  \includegraphics[scale=0.99]{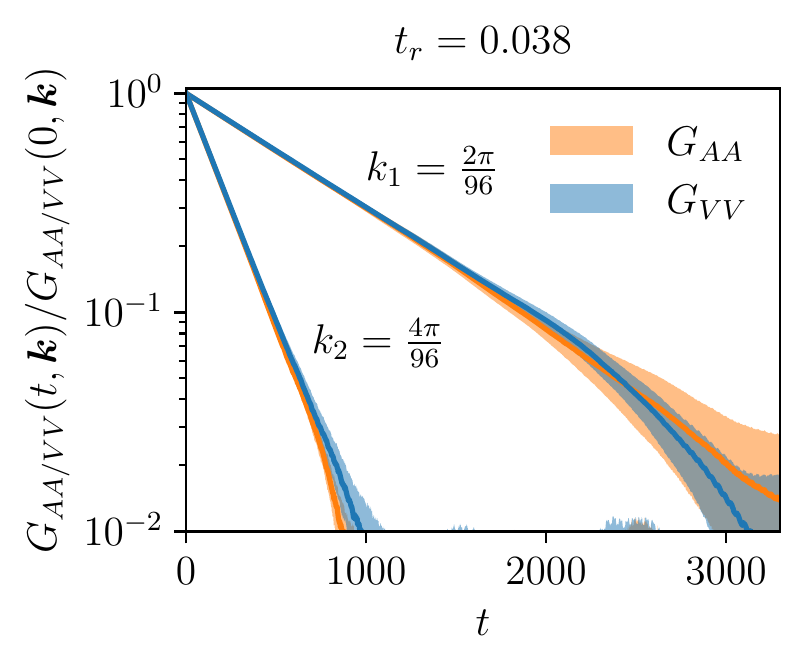}
  \includegraphics[scale=0.99]{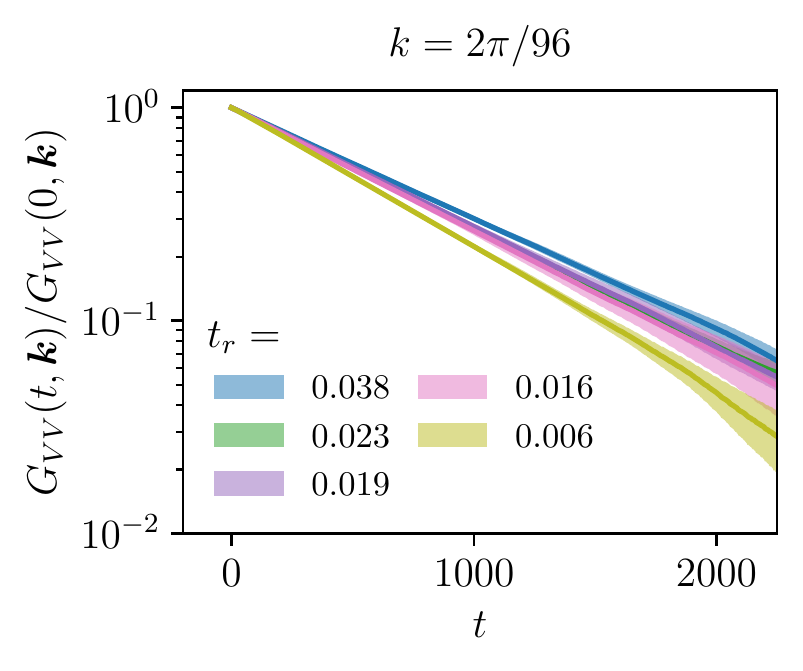}
  \caption{\textbf{Left:} Axial and vector charge correlators in the restored phase, for two different momenta. As expected, they decay exponentially (note the log scale) and the decay rate depends on $k^2$. \textbf{Right:} Vector correlator at a given finite $\boldsymbol{k}$ for different masses in the restored phase. A weak dependence of the diffusion coefficient on the mass is observed, pointing at a weak critical dependence for the diffusion coefficient; see Fig.~\ref{fig:dynamic_fit_restored} for a quantitative extraction.}
  \label{fig:AVrestored}
\end{figure}

In the restored phase, the $O(4)$ symmetry is unbroken and there is no distinction between the axial and vector channels. The order parameter simply dissipates towards equilibrium on a finite (if long) timescale. But, both the iso-vector and iso-axial vector charges are exactly conserved. As a result, the non-trivial dynamics at the longest wavelengths is fully encoded in the correlators of conserved charges at finite momenta. Indeed, for wavelengths long compared to the (diverging) correlation length $k\xi(T) \ll 1$, the diffusion equation remains a valid hydrodynamic description of the system:
\st
\label{eq:Diffusion}
        \partial_t n_{ab}  = D \nabla^2 n_{ab} \, .
\stp
Thus,  the relaxation rate of a diffusive mode of wavenumber $k$ is controlled by the diffusion coefficient $Dk^2$ leading to the correlators (see \App{app:mean_field}):
\st
G_{AA}(t,\k) = G_{VV}(t,\k)=T\chi_0 e^{- D k^2 t}\, .
\stp
The charge correlations in the simulation are illustrated on the
left-hand side of Fig.\ref{fig:AVrestored}, where we plot the axial correlator
(orange) and the vector correlator (blue), for the first two momenta, for a
given temperature. We clearly see that the two channels are degenerate. The $k$
dependence is consistent with the expected diffusive behavior.

In writing the diffusion equation in \Eq{eq:Diffusion} we have integrated out
the contributions to the currents from the order parameter, which
then leads the scaling behavior of the transport coefficient.  As discussed below,
the order parameter has a dynamical relaxation time of order $\tau
\propto \xi^{\zeta}$ with $\zeta=d/2$.  The diffusive modes have a relaxation
time of $1/Dk^2$, which is of order $\xi^2/D$ at the boundary of
applicability of the diffusion equation, $k \sim \xi^{-1}$. In the spirit
universality,  all modes of with wavelengths of order $\xi$ should  have a similar relaxation time, dictating  the expected scaling of the diffusion
coefficient, $D \propto \xi^{2-\zeta}$.

On the right-hand side of Fig.~\ref{fig:AVrestored}, we show the dependence of the vector correlator (we could have equivalently chosen the axial) on $t_r$. This very weak dependence seems at first to be in contradiction with the expected critical scaling of the diffusion coefficient $D\sim t_r^{(\zeta-2)\nu}$. As we will see more clearly in the next section, this apparent contradiction originates in the presence of a large regular constant contribution to $D$, and by the smallness of its non-universal critical amplitude.
Nevertheless, we will extract the critical contribution to $D$ in \Sect{sec:H0fitstau}.

\subsection{The critical line}
\label{sec:cline}

\begin{figure}
  \centering
  \subfloat{\includegraphics{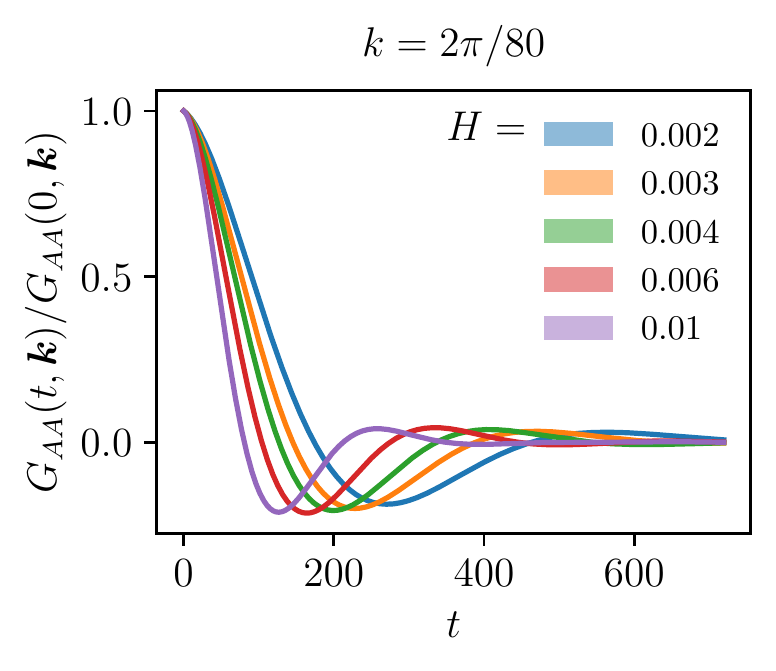}}\hfill
  \subfloat{\includegraphics{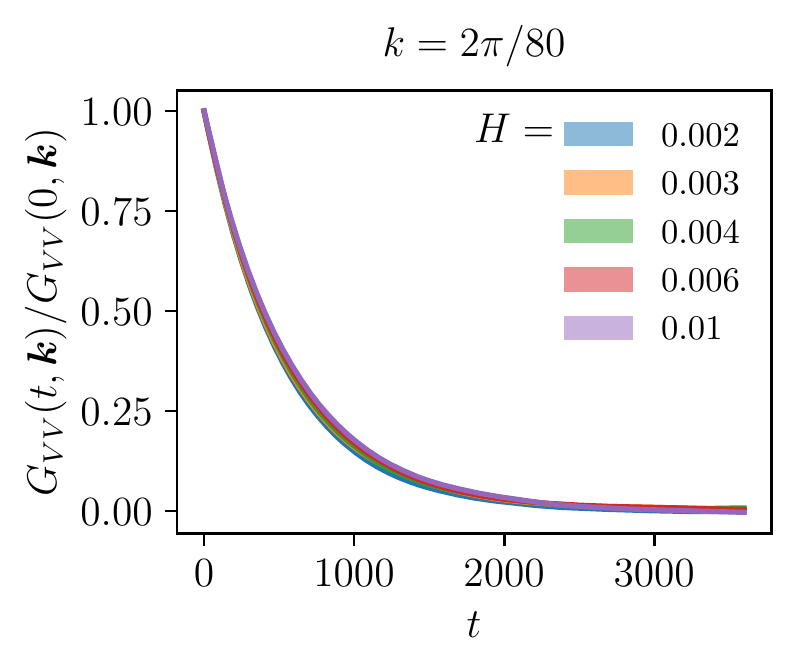}}\par
  \caption{\textbf{Left:} Axial correlators as a function of time at fixed momentum on the critical line ($t_r=0$) for different values of the magnetic field $H$. The axial charge is propagating but strongly damped and we see typical oscillations.
   \textbf{Right:} Vector correlator as a function of time at fixed momentum on the critical line for different magnetic fields. The vector charge remains diffusive. The symmetry is broken and the vector channel is no longer degenerate with the axial channel. As in the restored phase (see Fig.~\ref{fig:AVrestored}) we see a weak dependence on the temperature, indicating of a small critical amplitude which is extracted in Fig.~\ref{fig:dynamic_tau}.
   }
  \label{fig:critical_overview}
\end{figure}

We next study the critical line,  $m^2_0=m_c^2$, and study the effect of explicit symmetry breaking, $H\neq0$.
One of the qualitative outcomes of our previous work \cite{Florio:2021jlx} was to demonstrate the emergence of damped pion waves already on the critical line.
This qualitative observation was obtained from the $k=0$ modes of the axial current and the pions correlators.
The same qualitative observation holds for non-zero momentum $k$ and is observed in Fig.~\ref{fig:critical_overview}. At the level of the charge correlators, the magnetic field lifts the degeneracy between the axial and vector channels. The axial correlator then displays characteristic oscillations associated with axial charge propagation in the form of damped pion waves. The vector channel on the other hand remains purely diffusive. As in the restored phase, we already observe that the critical behavior of the vector diffusion constant is weak.

\subsection{Broken phase with $H=0$}
\label{sec:broken}

Below the critical point the longest wavelength modes are characterized by
massless Goldstone bosons, which describe phase fluctuations  of the chiral
condensate\footnote{For clarity we will restore the temperature in this section.}.  Following  \cite{Son:2002ci,Son:2001ff}, these fluctuations  are parametrized by
an angle $\vec{\varphi}(t,\x)$ with
$\phi_a = (\bar \sigma, \bar\sigma \vec{\varphi}(t,\x))$.  
We are assuming that the chiral condensate is nearly aligned with the pole of the three sphere (the coset manifold) and takes the form
\st
    \frac{1}{V} \int_{\x}d^3x \, \phi_a(t,\x) = \bar\sigma \, n_a \,  ,
\stp
where $n_a$ is a unit vector that is approximately parametrized by $n_a \simeq (1,\vec{\varphi}_0)$  to first order in the angular deviation.
When the wavelength of $\varphi$ is long compared to the
correlation length $\xi\equiv m_{\sigma}^{-1}$, a hydrodynamic description of
$\varphi$ is valid,  and the  fluctuations of modes with wavelength
order $m_\sigma^{-1}$ simply determine the critical behavior of the parameters of the hydrodynamic theory.

The free energy associated with the phase and charge fluctuations in the hydrodynamic  approximation is
\st
\label{eq:varphifree}
 \mathcal  H[\varphi,n] =\int {\rm d}^3x \, \frac{\vec{n}_V^2}{2\chi_0} + \frac{\vec{n}_A^2}{2\chi_0} +   \tfrac{1}{2}  f^2(T)  \,  \partial_i \vec{\varphi} \cdot  \partial^i \vec{\varphi} +  \tfrac{1}{2} f^2 m^2\varphi^2  \, ,
\stp
where we have included a mass term, which is relevant only when the magnetic field is non-zero and is determined by the Gell-Mann, Oakes, Renner relation
\st
      f^2 m^2 = H \bar{\sigma} \, .
\stp
We will set $H=m=0$ for the remainder of this section.
The resulting hydrodynamic equations of motion for the pions and axial charge take the form\footnote{
   In \cite{Grossi:2020ezz} the dissipative coefficient $\Gamma/f^2$ was called $\zeta^{(2)}$. In mean field theory the hydrodynamic parameter $\Gamma$ equals the bare parameter $\Gamma_0$~\cite{Grossi:2021gqi}.  A notable feature 
   of these equations when the symmetry is 
   explicitly broken is that the same coefficient $\Gamma/f^2$ controls both the 
   damping rate at finite momentum and the mass term, see \Eq{eq:hydroeqns2}. This is a general result dictated by entropy considerations~\cite{Rajagopal:1992qz,Grossi:2020ezz,Armas:2021vku} and related locality arguments~\cite{Delacretaz:2021qqu}. This became widely recognized through a sequence of explicit holographic computations in different contexts with explicitly broken symmetries~\cite{Ammon:2019wci,Amoretti:2018tzw,Donos:2021pkk,Ammon:2021pyz} -- for a review see~\cite{Baggioli:2022pyb}. The pion dynamics in a holographic model of the phase transition is described in \cite{Cao:2022csq} and can be profitably compared to mean field theory results of \cite{Grossi:2021gqi}.  
}
   \begin{subequations}
\label{eq:hydroeqns}
\begin{align}
\partial_t \vec{\varphi}  - \frac{\vec{n}_A}{\chi} =& \frac{\Gamma }{f^2} \,  \nabla \cdot (f^2  \nabla \vec{\varphi})  \, ,  \label{eq:varphieqns} \\
\partial_t  \vec{n}_A -  \nabla (f^2 \nabla \vec{\varphi})  =& D \nabla^2 \vec{n}_A  \, ,
\end{align}
which reflects from the Poisson bracket between the phase and the charge, $\{\varphi_{s},n_{As'}\}=\delta_{ss'}$. The vector charge remains diffusive
\st
\partial_t  \vec{n}_V  = D \nabla^2 \vec{n}_V  \, .
\stp
\end{subequations}

The hydrodynamic system is solved in \App{app:mean_field} and the resulting pion eigen-waves have dispersion relations
\st
   \omega(k) - \frac{i}{2} \Gamma(k)= \pm v k  -  \frac{i}{2} D_A k^2  \, ,
\stp
where
\st
\label{eq:vandDa}
    v^2 \equiv \frac{f^2}{\chi_0}  \quad \mbox{and} \quad D_A \equiv \Gamma + D  \, .
\stp
The  form of the  correlation function from the  hydrodynamic theory is determined  in \App{app:mean_field}
\st
\label{eq:Gvphivphidynamic}
   G_{\varphi\varphi}(t,\k) = \frac{T}{f^2 k^2} e^{-\frac{1}{2} D_A k^2 t} \left[ \cos(v k t) + \frac{1}{2 v k} (D - \Gamma) k^2 \sin(v k t) \right] \, .
\stp

At large wavelengths the pion effective theory determines the static and dynamic correlation functions. The static correlation function for $k\ll m_\sigma$ can be read off from the free energy in \eqref{eq:varphifree} and is singular in the massless limit:
\st
           G_{\pi\pi}(0,\k) =\bar \sigma^2 G_{\varphi\varphi}(0, \k) =  \bar\sigma^2\, \frac{T}{f^2 k^2}\, ,   \qquad k \ll m_\sigma \, .
\stp
By contrast, the static fluctuations of the $\sigma$
are bounded (if diverging) and determined by the static susceptibility of the
$O(4)$ critical point, which scales as 
\st
G_{\sigma\sigma}(0,\k)  \propto m_\sigma^{\eta-2} \qquad k  \ll m_\sigma  \, .
\stp
The $\sigma$ and $\vec{\pi}$ are part of an $O(4)$ multiplet and their correlators must be the same order of magnitude at the boundary of applicability of the pion EFT,  $k\sim m_\sigma$.
Since the vev scales  as $\bar \sigma^2 \propto m_\sigma^{(d-2 + \eta)}$, the pion constant must scale as
$f^2 \propto m_\sigma^{(d-2)}$~\cite{Hasenfratz:1989pk}. We have anticipated this scaling in \Fig{fig:critical_overview} where $T/2f^2$ (with $T=1$  in our adopted units) served as a definition of the correlation length in the broken phase.

\begin{figure}
  \centering
  \subfloat{\includegraphics{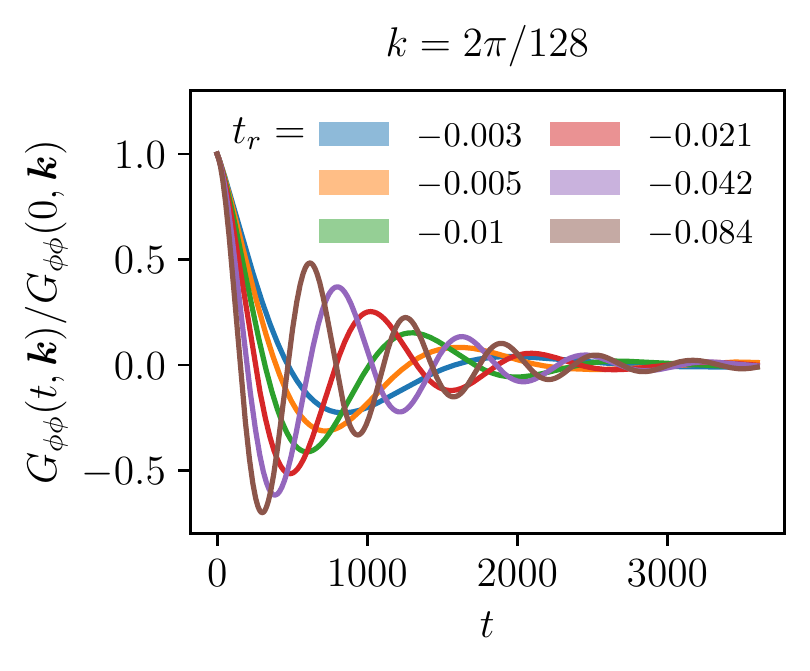}}
  \subfloat{\includegraphics{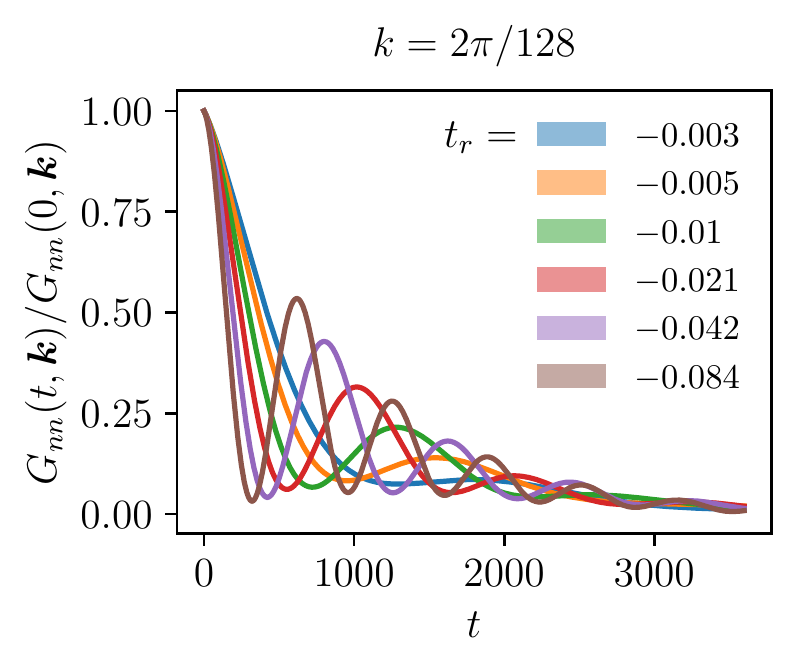}}
  \caption{
  \textbf{Left:}
Normalized $G_{\phi\phi}$ correlator at $k=2\pi/128$ as function of time  for different reduced temperatures $t_r$ in the broken phase. We see that this correlator is sensitive to the Goldstone modes.
  \textbf{Right:} Normalized
   $G_{nn}$ correlator at $k=2\pi/128$ as function of time for different reduced temperatures $t_r$ in the broken phase. This correlator is also sensitive to the Goldstone dynamics.}
  \label{fig:show_H=0}
\end{figure}

Because of these scalings, the long-wavelength behavior of the static and dynamic correlation functions of $\phi_a$ are dominated by the phase fluctuations. Specifically,  the static correlation function reads
\begin{align}
\label{eq:Gphiphistatic}
G_{\phi\phi}(0,\k)  = \frac{1}{4} \left( G_{\sigma\sigma}(0,\k) + 3 G_{\pi\pi}(0,\k) \right) 
\simeq  \frac{3}{4} \, \bar\sigma^2 \, \frac{T}{f^2 k^2}\, ,     \qquad k \ll m_\sigma \, ,
\end{align}
up to corrections of order $(k/m_\sigma)^2$. Additional finite volume corrections to this formula are given in \App{app:staticanalysis} and are suppressed by $1/f^2L$~\cite{Hasenfratz:1989pk}.  Phase fluctuations also determine the dynamic correlations
\st
       G_{\phi\phi}(t,\k) \simeq \frac{3}{4} \, \bar \sigma^2 G_{\varphi\varphi}(t,\k) \, ,
\stp
and Fig.~\ref{fig:show_H=0} (Left) exhibits the $\phi\phi$ correlation function, which shows characteristic pion oscillations of \eqref{eq:Gvphivphidynamic}  become increasingly damped near the critical point.
The critical scaling of $v$ and $D_A$ are extracted numerically in the next section and are ultimately summarized in \Fig{fig:v_axial_dynamic}.

In a finite volume simulation the orientation of the chiral condensate is not fixed, but diffuses in time due to  thermal noise.
Indeed,
 a stochastic variable $\xi_\varphi$ with variance 
\st
           \llangle \xi_{\varphi s}(t,\x) \xi_{\varphi s'}(t',\x') \rrangle =  \frac{2 T\Gamma}{f^2} \delta_{ss'} \delta(t -t') \delta(\x -\x') \, ,
\stp
should be added to \eqref{eq:varphieqns}. 
Integrating this equation over the volume shows that 
the angular zero mode   
$\vec{\varphi}_0(t) \equiv \int_\x \vec{\varphi}(t,\x)/V$ 
has a variance  which increases in time  
\st
\label{eq:vevvariance}
  2 \, \mathfrak{D}\,  t =  \frac{1}{3} \llangle \vec{\varphi}_0(t) \cdot \vec\varphi_0(t)  \rrangle = 2 \left(\frac{T\Gamma }{f^2 V}\right) t \, .
\stp
This result (which is presented in detail in \App{sec:FokkerPlanck}) relates the vev diffusion coefficient  to the pion kinetic coefficients, $\mathfrak D = T\Gamma/f^2 V$. 
The diffusion rate decreases with increasing volume and is suppressed relative to the pion damping rate,  $\Gamma k^2 \sim \Gamma/L^2$,  by one power of length, $T/f^2 L\propto  1/m_\sigma L$. 
When the magnetic field is non-zero 
the Fokker-Planck equation describing the motion of the chiral condensate on the three sphere  features a rich interplay between Hamiltonian and diffusive dynamics, which is developed in \App{sec:FokkerPlanck}. 

Turning to the conserved charges, the correlation function of the vector and axial-vector charges in the broken phase are intrinsically different. The axial charge is coupled through in equations of motion to the pions and its correlation function reflects this coupling (see \App{app:mean_field})
\st
\label{eq:GAAeqn}
         G_{AA}(t,\k) =
      T \chi_0 e^{-\tfrac{1}{2} D_A k^2 t} \left[ \cos(v k t) - \frac{1}{2vk} (D -\Gamma)k^2 \sin( v k t) \right] \, .
\stp
The vector charge continues to obey the diffusion equation with corresponding response functions
\st
         G_{VV}(t,\k) = T\chi_0 e^{-Dk^2 t}   \, .
\stp
Separating the vector and axial vector response in the absence of
explicit symmetry breaking is challenging, since the
condensate forms in an arbitrary direction. Moreover, because of finite volume, this direction is not completely fixed, but slowly wanders along the coset manifold.  This can be addressed in future work. For now we have simply
computed the combined correlator $G_{nn}(t,\k)$
\st
     G_{nn}(t,\k) = \frac{1}{6} \left[3 G_{AA}(t,\k) + 3 G_{VV}(t,\k) \right]  \, ,
\stp
which is exhibited in \Fig{fig:show_H=0} (Right).
The result curves show a mix of exponential decay and oscillating pion waves.


It is striking how the hydrodynamic EFT and the pattern of chiral symmetry break essentially dictates the scaling dynamical response of ``Model G"~\cite{Son:2002ci,Son:2001ff}. A summary of the reasoning proves an overview of the next section.
\begin{enumerate}[label=(\roman*)]
   \item  The characteristic frequency of the pion waves below the critical point is of order $\omega(k) \sim v k$ where $v = \sqrt{f^2/\chi_0} \propto m_{\sigma}^{(d-2)/2}$.   The scaling of the velocity will be extracted from the real time correlations of $\phi$, \Eq{eq:Gvphivphidynamic}, and from its static behavior, \Eq{eq:Gphiphistatic}, and the results are summarized in \Fig{fig:v_axial_dynamic} (Right).
   \item
  At the boundary of applicability of the pion EFT $k\sim m_{\sigma}$, the pion frequency is $\omega(k) \sim v k \sim m_{\sigma}^{d/2}$.
  But, since the pions are part of an $O(4)$ multiplet, the order parameter must inherit this dynamical timescale, $\tau \sim m_{\sigma}^{-d/2}$, setting the dynamical critical exponent of Model G, $\zeta=d/2$. $\tau$ will be extracted in the restored phase, \Fig{fig:dynamic_fit_restored} (Left),  and on the critical line, \Fig{fig:dynamic_tau} (Left).
  \item Close to the critical point and for $k \lesssim m_\sigma$, the real and imaginary parts of $\omega(k)$ should be commensurate if the critical point is characterized by a single time scale. This reasoning dictates the scaling of the pion damping coefficient, $D_A \propto m_{\sigma}^{d/2 - 2}$.  The scaling of $D_{A}$ is exhibited in \Fig{fig:v_axial_dynamic} (Left).
  \item Since the axial and vector charges must become degenerate above
     $T_c$, the diffusion coefficient must have the same scaling as $D_A$, leading to the scaling,  $D\sim m_{\sigma}^{d/2 -2}$. The scaling of $D$ is exhibited in the restored phase, \Fig{fig:dynamic_fit_restored} (Right),  and on the critical line, \Fig{fig:dynamic_tau} (Right).
     \end{enumerate}

Finally, the vev diffusion coefficient $\mathfrak{D}$ is also consistent with these scalings. Taking the scaling of $f^2 \propto m_{\sigma}^{d-2}$ and $\Gamma \propto m_\sigma^{d/2 -2 }$ we see the angular variance given in \eqref{eq:vevvariance} grows as
\st
            2 \, \mathfrak{D} \, t \propto  \frac{t}{(m_\sigma L^2)^{d/2}}   \, .
\stp
At the critical point  where $m_{\sigma}L \rightarrow  1$, the angular variance increases as
\st
         2\, \mathfrak{D} \, t \rightarrow \frac{t}{L^{d/2}} \, ,
\stp
which is again consistent with a universal dynamical critical timescale of $\tau \propto L^{\zeta}$ with a common critical exponent of $\zeta = d/2$.

\section{Relaxation time and finite momenta dynamics}

\subsection{$H=0$, restored phase:  scaling of the correlation time}
\label{sec:H0fitstau}


We first determine the order parameter relaxation time $\tau$  from the two-point function $G_{\phi\phi}(t,\k)$.
Specifically, we define $\tau$ as the autocorrelation time of $G_{\phi\phi}$ at zero momentum $G_{\phi\phi}(t)\equiv G_{\phi\phi}(t,{\bm 0}) $
\begin{equation}
  \tau \equiv \int_0^{\infty} \mathrm{d}t \, \frac{G_{\phi\phi}(t)}{G_{\phi\phi}(0)} \ ,
\end{equation}
which is motivated by an exponential decay, $G_{\phi\phi} \propto e^{-t/\tau}$
(in practice, we cutoff the integral at some $T_{max}$ and make sure the results are independent of this technicality).
Qualitatively,  the decorrelation of $G_{\phi\phi}(t)$ is characterized by a single timescale $\tau$ without significant structure. 
\begin{figure}
  \centering
  \subfloat{\includegraphics[scale=1]{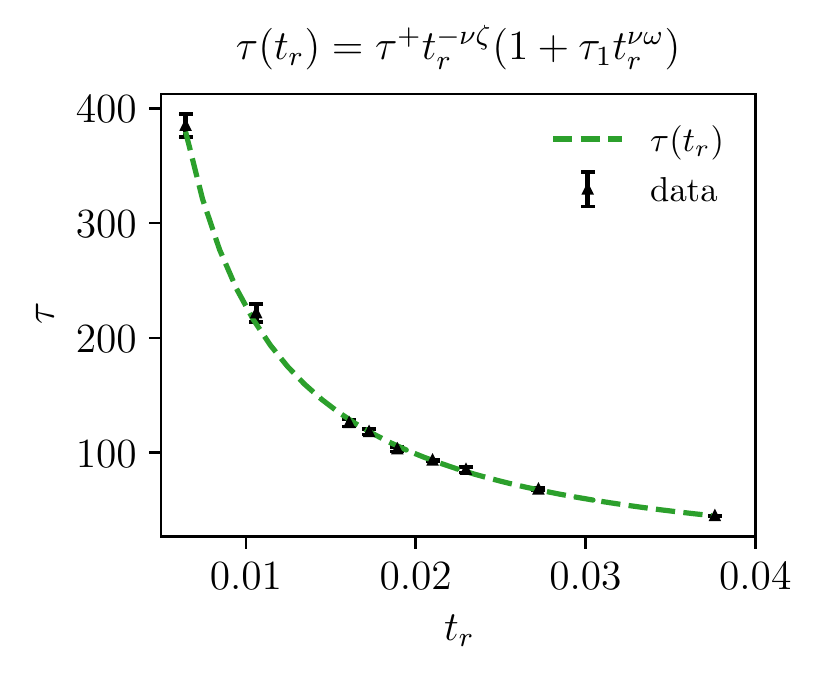}}
  \subfloat{\includegraphics[scale=1]{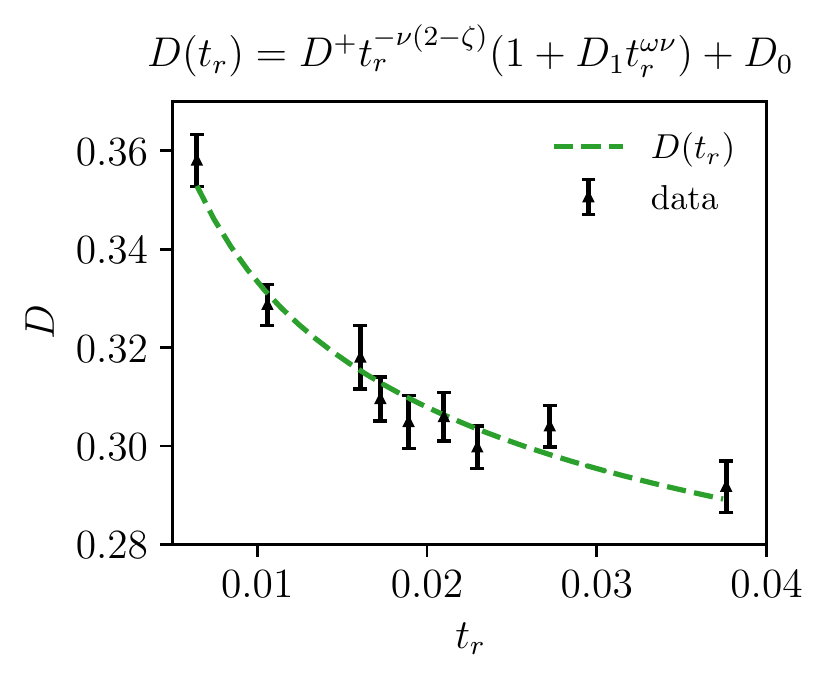}}
  \caption{\textbf{Left}:  Dynamical correlation time as a function of $t_r$ in the restored phase. The fit parameters are   $\tau^+ = 1.570 \pm 0.037, \ \ \tau_1 = -1.49 \pm 0.15$ with $\chi^2/\mathrm{dof}=5.74/7$.
  \textbf{Right}: Vector diffusion coefficient as a function of $t_r$. Note that despite its smallness, we clearly observe a critical dependence of $D$. The resulting fit parameters are $ D^+=0.0190 \pm 0.0013 , D_1=-1.7 \pm 0.55$ with $\chi^2/\mathrm{dof}=5.67/7$. }
  \label{fig:dynamic_fit_restored}
\end{figure}
From scaling, we expect the correlation time to behave as
\begin{equation}
  \tau(t_r)=\tau^+ t_r^{-\nu\zeta}(1+\tau_1t^{\nu\omega}) \, ,
\end{equation}
with the critical exponents $\nu$, $\omega$, and  $\zeta$ given in Table~\ref{tab:num_values_fixed}  in App.~\ref{Sec:StaticMeasurements}.
The resulting fit (with the exponents fixed) is shown in Fig.~\ref{fig:dynamic_fit_restored} and we find
\begin{align}
  \tau^+ &= 1.570 \pm 0.037, \\
  \tau_1 &= -1.49 \pm 0.15 \ .
\end{align}
The growth of $\tau$ near the $T_c$ is nicely consistent with $\zeta=d/2$.

The diffusion coefficient $D$ of the $O(4)$ charges is also expected to show dynamical critical
scaling, but since the charges are conserved, the coefficient has to be extracted from  finite momentum correlators. In particular,
 we expect the vector correlator in the restored phase to behave as
\begin{equation}
G_{VV}(t,\k) = T\chi_0 e^{-D k^2 t}  \ ,
\end{equation}
as discussed in the previous section.
Similar to  the order parameter case, we have the relation
\begin{equation}
 D \equiv \frac{1}{k^2}\left(\int_0^\infty \mathrm{d}t \, \frac{G_{VV}(t,\k)}{G_{VV}(0, \bm 0)} \right)^{-1} \ ,
\end{equation}
which we use in practice to extract $D$ from our data.  It is important that the correlator be well described by an exponential, which is clearly seen in \Fig{fig:AVrestored}.

Again, we introduce a cutoff $T_{max}$ and make sure the results are independent of this choice. As discussed above,  we expect $Dk^2$ to be comparable $1/{\tau} \sim \xi^\zeta$ at the boundary of applicability of the diffusive description, $k \lesssim \xi$.  This means that the diffusion coefficient should scale with the correlation length as,  $D\propto \xi^{2-\zeta}$.

The critical behavior of $D$ appears to be weak and its regular part cannot be neglected.
 To take this into account, we add a constant $D_0$ to our fit
\begin{equation}
  D(t_r) = D^+ t_r^{-\nu(2-\zeta)}(1+D_1t_r^{\omega\nu}) + D_0 \ . \label{eq:fitDrestored}
\end{equation}
In an abuse of notation, $D_0$ here is the renormalized diffusion coefficient far above $T_c$, which we determine independently by running our simulations without the coupling to the order parameter, leading to
  \begin{equation}
    D_0 \approx 0.2425(5) \ .
  \end{equation}
  This parameter is different from the bare lattice parameter (also called $D_0=1/3$) which controls the Metropolis updates of the charge at the scale of the lattice spacing (see \cite{Florio:2021jlx} in Appendix A.c for further details).
Fixing $D_0$ and fitting $D^+$ and $D_1$ in  \eqref{eq:fitDrestored}, we get  the following determination of the non-universal amplitudes
 \begin{align}
  D^+&=0.0190 \pm 0.0013\, , \\
  D_1&=-1.7 \pm 0.55 \ .
\end{align}
$D^+$ is small as was anticipated based on a one loop calculation using mean field propagators~\cite{Grossi:2021gqi}.

Using the non-universal amplitude $\xi^+$ from the correlation length $\xi\sim \xi^+ t_r^{-\nu}$ determined in \App{app:corrlength_unbroken}, we can combine these results  into a universal dynamical amplitude ratio
\begin{align}
\QD^+ \equiv  \frac{D^+ \tau^+}{\xi^{+ 2}} = 0.148\pm0.011 \ ,
\label{eq:univratiorestored}
\end{align}
which numerically encodes the coupling between the charge diffusion  and the order parameter relaxation  at criticality.
The universal dynamical ratio presented here is new and will prove useful to any future study which realizes the critical dynamics of Model G.

\subsection{The critical line}
\begin{figure}
  \centering
 	\subfloat{\includegraphics{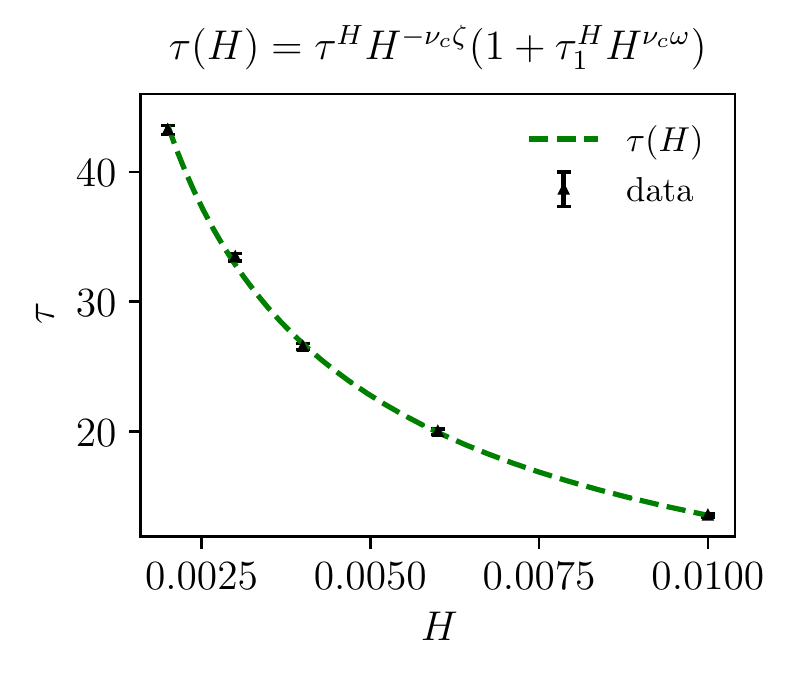}}
	\subfloat{\includegraphics{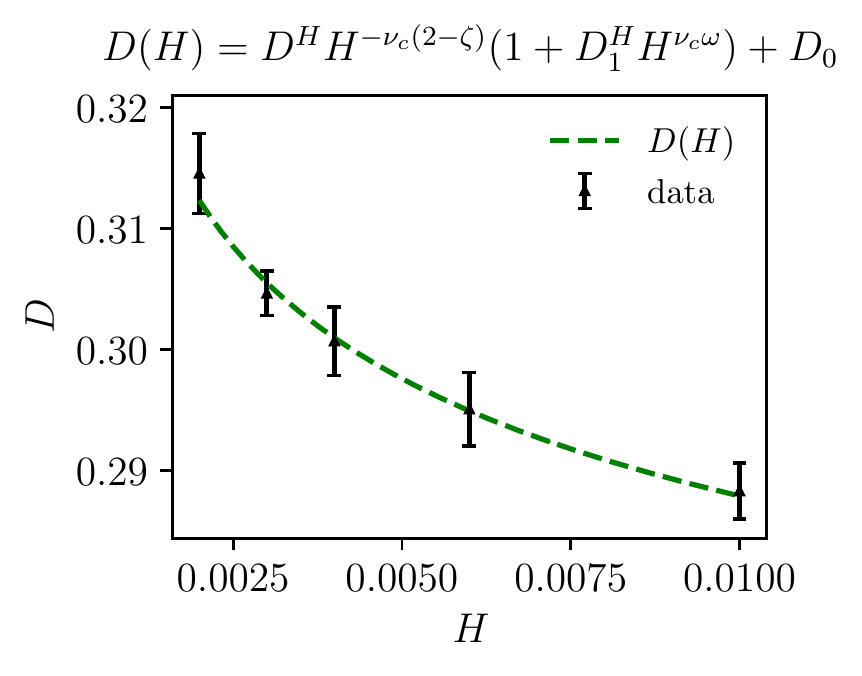}}
  \caption{\textbf{Left:} Relaxation time for the $\sigma$ field, $\tau(H)$,  as function of the magnetic field $H$ on the critical line, $m^2_0=m^2_c $.
The data points are fit with $
\tau(H) = \tau^H H^{-\nu_c \zeta} (1 + \tau_1 H^{\nu_c \omega})
$. The results are
$\tau^{H} =1.312 \pm 0.022$ and  $\tau_1^H=-1.45 \pm 0.06$ with $\chi^2/\mathrm{dof}=6.07/3$.
 \textbf{Right:} Diffusion coefficient as function of the magnetic field $H$ on the critical line,  $m^2_0=m^2_c $.
The data points are fit with $
D(H) = D^{H} H^{-\nu_c \zeta} (1 + D_1^H H^{\nu_c \omega}) +D_0
$. The results are
$D^{H} =0.023 \pm 0.002$ and  $D_1^H=-1.02 \pm 0.42$ and $D_0=0.2425$ is kept fixed with $\chi^2/\mathrm{dof}=0.72/ 3$.
 }
 \label{fig:dynamic_tau}
\end{figure}

We next study the dynamics on  the critical line,   $m^2_0 = m_c^2$ and $H \neq 0$. As in the previous section we define the order parameter relaxation time
$\tau$ from the correlation  of the $\sigma$ field
\begin{equation}
\tau \equiv \int_{0}^{\infty} \dd t \frac{G_{\sigma \sigma }(t,\bm 0)}{G_{\sigma \sigma }(0,\bm 0)} \ .
\end{equation}
%
The results for $\tau$ as a function of the magnetic field are shown in \Fig{fig:dynamic_tau}, and
are fit with the expected scaling form
\begin{equation}
\tau(H) = \tau^H H^{-\nu_c \zeta} (1 + \tau_1 H^{\nu_c \omega}) \ ,
\end{equation}
with fixed exponents $\nu_c$, $\omega$,  and $\zeta$ given in Table~\ref{tab:num_values_fixed} of \App{Sec:StaticMeasurements}, and fitted parameters
\begin{align}
  \tau^{H} &=1.312 \pm 0.022\, ,\\
  \tau_1^H &=-1.45 \pm 0.06 \ .
\end{align}
As in the restored phase,  the subleading term with exponent $\omega$ is  necessary to have a reasonable fit compatible with the expected dynamical exponent, $\zeta=d/2$.

Similarly, the diffusion coefficient is defined in the same way as in the restored phase
\begin{equation}
D \equiv \frac{1}{k^2 } \left(\int_{0}^{\infty} \mathrm{d}t\, \frac{G_{VV}(t,\k)}{G_{VV} (0,\k)} \right)^{-1} ,\text{ with } |\k|= \frac{2\pi}{L},
\end{equation}
and we adopt an analogous strategy and functional form to fit the dependence of $D$ on the correlation length
\begin{equation}
D(H)= D^{H} H^{-\nu_c(2-\zeta)}(1+ D_{1}^H H^{\nu_c \omega} ) +D_0 \, .
\end{equation}
As in the previous section, $D_0=0.2425$ is the regular part of the diffusion coefficient and is extracted from our diffusion only runs.
The fit results are shown in Fig.~\ref{fig:dynamic_tau} and the fit parameters are
\begin{align}
  D^{H} &=0.023 \pm 0.002 \, , \\
  D_1^H&=-1.02 \pm 0.42 \ .
\end{align}
The singular part of the diffusion coefficient diverges next to the critical point with the expected
exponent $H^{-\nu_c(2-\zeta)}$ and the subleading correction improves the quality of the fit.  Including the regular part of the diffusion coefficient $D_0$ is crucial to extracting the scaling of $D$ with $H$, since the critical amplitude $D^{H}$ is small, as anticipated in the one loop computation in \cite{Grossi:2021gqi}.

The results for the order parameter and the diffusion coefficient can be concisely recast as a universal amplitude ratio
\begin{align}
  \QD^{H} \equiv\frac{D^H \tau^H}{\xi^{H 2}_L} = 0.151\pm 0.022  \, ,
\end{align}
where $\xi_L^H$ is the amplitude associated with the longitudinal correlation length  and is determined in \App{app:corrlengthcrit}. Note that within numerical errors the dynamical ratio on the critical line matches the one in the restored phase \eqref{eq:univratiorestored}.

\subsection{$H=0$ and the broken phase}

Our aim in this section is to extract the matching coefficients of the pion hydrodynamic EFT 
by studying the momentum dependence of the correlators and analyzing the pion dispersion curve. To this end, we study $G_{\phi\phi}$ for various $t_r$ and $L$ and analyze the two lowest non-zero momentum  modes, $k=2\pi/L$ and  $k=4\pi/L$. We then perform some fits  using the hydro prediction \eqref{eq:Gvphivphidynamic} and extract the parameters $\omega(k)$ and $\Gamma(k)$. $\omega(k)$ and $\Gamma(k)$ are subsequently extrapolated to infinite volume using the forms
\begin{subequations}
\label{eq:fittingform}
\begin{align}
\Gamma(k)&= D_A k^2(1+d_1 /L),\\
\omega(k)&=v k + v_2 k^2 \ ,
\end{align}
\end{subequations}
with $D_A$, $v$, $d_1$, and $v_2$  as parameters.
As discussed in \Sect{sec:broken}, finite volume corrections to the dynamics of the Goldstone modes  scale inversely with the linear extent of the lattice  $\sim 1/f^2L$ and are therefore important to control.
Including $d_1$ and $v_1$ as fit parameters captures this characteristic volume dependence and proved crucial to reliably extracting $D_A$ and $v$ from this analysis.
The fitting procedure, together with other systematic checks,  are  presented in detail in App.~\ref{app:axial_channel}.

In the static case, a finite volume pion EFT
was developed  many years ago by Hasenfratz and Leutwyler
and can be used to determine the first $1/f^2L$ correction to the chiral condensate analytically~\cite{Hasenfratz:1989pk}. We used their expansion in \App{app:staticanalysis} to determine  velocity $v^2=f^2/\chi_0$ and, as we show below, the result agrees with the dynamic fit using \eqref{eq:fittingform}.  In \Sect{sec:broken} we have taken the first steps in developing a real time finite volume analog of their work by deriving the vev diffusion coefficient in \eqref{eq:vevvariance} and a corresponding forced Fokker-Planck equation on the coset manifold in \app{sec:FokkerPlanck}. But,  a more complete analysis is left for future work.

\begin{figure}
  \centering
    \subfloat{\includegraphics[scale=0.98]{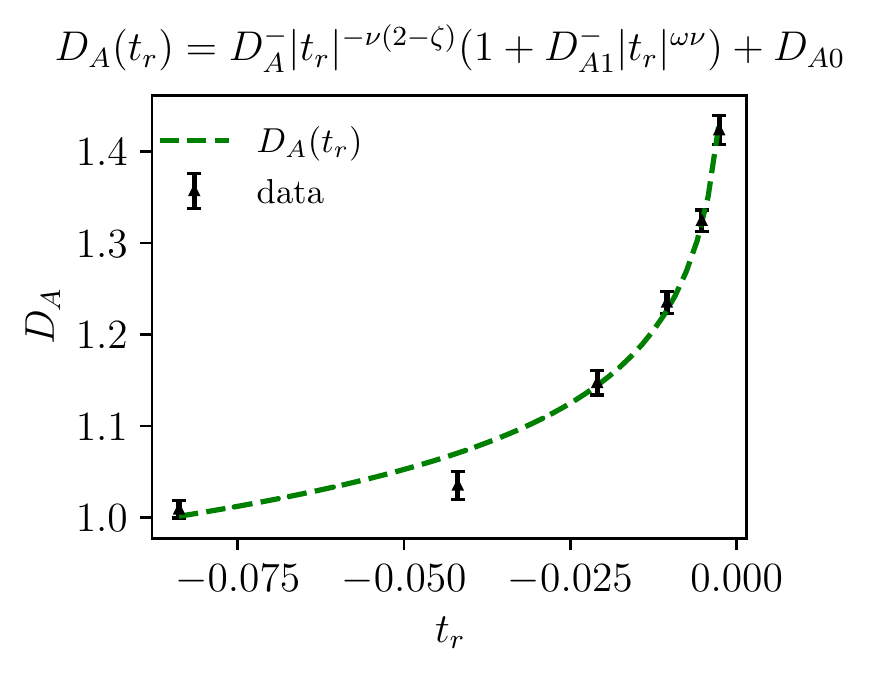}}
    \subfloat{\includegraphics[scale=0.98]{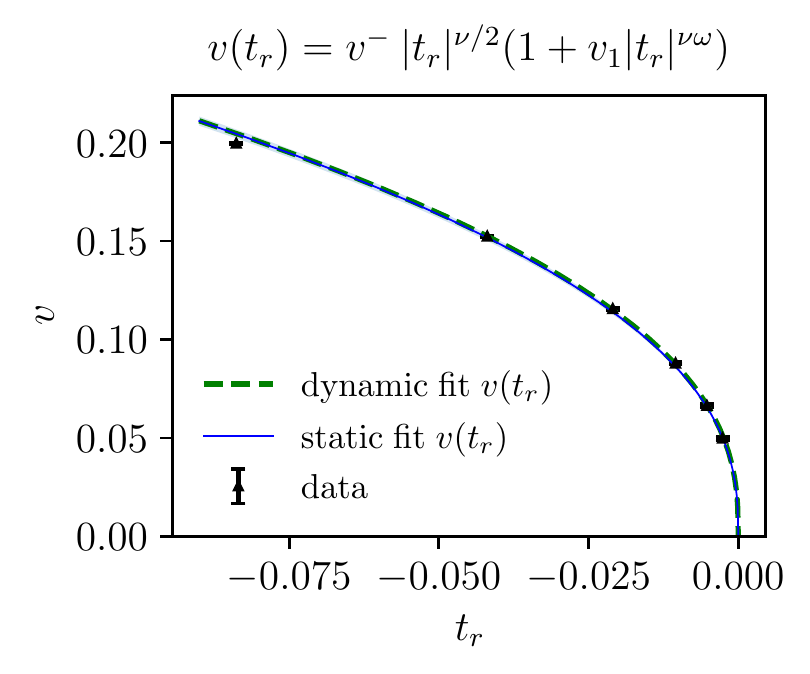}}
 	  \caption{
	 \textbf{Left}:
	 $D_A=\Gamma+D$ axial diffusion coefficient as a function of the reduced temperature $t_r$.
	  The dashed line is a fit to the scaling form
	 $D_A = D_A^- t_r^{-\nu(2-\zeta)}(1+D_{A1}t_r^{\omega\nu}) + D_{A0}  $.
	 The fit results are $D_A^-= 0.044\pm0.009$, $D_{A1}= 1.19\pm 0.13$ and $D_{A0}= -11\pm 6$.
	 The $\chi^2/\mathrm{dof}=6.9/3$.
   \textbf{Right}:
   Velocity of the dispersion relation as a function of $t_r$.
   The data points are the value extracted from the time dependence of the correlation functions.
   The dashed (green) line is the fit to the data with the scaling form
  $v =v^{-} \:t_r^{\beta/2} (1 + v_1 t_r^{\nu\omega }  )$.
  The fit results are $v^{-}= 0.462\pm0.003$ and $v_1= 0.325\pm 0.03$.
  The $\chi^2/4=15.9/4$.  The continuous (blue) line is described in  \App{app:staticanalysis}
  and in the text and is obtained  by determining $v^2 = f^2/\chi_0$ from static correlation measurements.
	 }
  \label{fig:phi_diffusion}
   \label{fig:v_axial_dynamic}
\end{figure}

Knowing $v$ and $D_A$, we can proceed further and study their  scaling with temperature. We start with a study of the critical behavior of $D_A$, shown on the left hand side of Fig.~\ref{fig:v_axial_dynamic}.
We use the following scaling ansatz
\begin{equation}
D_A(t_r) = D_A^{-}|t_r|^{-\nu(2-d)}(1+ D_{A1}^- |t_r|^{\omega \nu}) +D_{A0} \ ,
\end{equation}
and obtain the following fit parameters
 \begin{align}
 D_A^{-}&= 0.044 \pm 0.009,\\
D_{A1}^-&= 1.19 \pm 0.13,\\
D_{A0}&= -11\pm6.
\end{align}
with a $\chi^2 /\text{dof}$ of $6.3/3$, which is mostly driven by the large fluctuation around $t_r\sim-0.004$.
The exponents $\nu$, $\zeta$ and the subleading $\omega$  were held fixed.
As in all the previous cases, including a subleading correction and a regular part is necessary to obtain a precise fit. We did not find a way to independently fix the regular part for this analysis. This leads to a  degeneracy between $D_{A0}$ and $D_{A1}^-$, resulting in large errors for both of these parameters. This problem does not affect the extracted value of the leading amplitude, $D_A^{-}$.

The Goldstone velocity $v$ is shown on the right hand side of Fig.~\ref{fig:v_axial_dynamic}.
The data are fitted with the expected scaling form,
\begin{equation}
v (t_r)=v^{-} \:t_r^{\nu/2} (1 +v_1 t_r^{\nu\omega }  ) \ ,
\end{equation}
with $\nu$ and $\omega$ fixed, yielding
\begin{align}
v^{-}&= 0.462\pm0.003 \, , \\
v_1&= 0.325\pm 0.03 \ .
\end{align}
The $\chi^2$ is $15.9$  with $4$ degrees of freedom. The relatively large value of the $\chi^2$ is due to the very high precision of our data. It is driven by the points farther away from the critical point and probably reflects the fact that our data are sensitive to sub-subleading corrections.  We also show the determination of $v=\sqrt{f^2/\chi_0}$ obtained from the static analysis in App.~\ref{app:staticanalysis}. The analysis makes use of \eqref{eq:Gphiphistatic} but exploits the known finite volume analysis of Hasenfratz and Leutwyler~\cite{Hasenfratz:1989pk}.  The agreement between the static analysis and the fits to the real time correlation functions is remarkable and is a highly non-trivial confirmation of the low-energy EFT.

Ideally, at this point we would complete this study by determining the critical amplitude of the diffusion coefficient in the broken phase. However,
because the amplitude is small, and because finite volume corrections are only suppressed by $1/f^2L$, we were unable to unambiguously extract this amplitude with the current dataset and our current understanding of finite volume corrections. We can clearly see the diffusive behavior,  but the coefficient is constant within uncertainty.


\section{Discussion}

In this work, we performed an extensive study of the finite momentum
dynamics of ``Model G", the critical dynamical model corresponding to
two-flavor QCD in the chiral limit. Across all different phases of the theory, we exhibited striking
agreement between theoretical predictions and our full-fledged
numerical simulations. In particular, although the critical amplitude
is small, we showed that the critical contribution to the
vector diffusion coefficient $D$ scales with the correct
dynamical critical exponent, both in the restored phase and along the
critical line. After additionally extracting  the critical relaxation
time $\tau$ of the order parameter (which also scales appropriately), we  recast
these new results as  universal dynamical amplitude ratios $\QD^+$, $\QD^{H}$, which reflect the coupling between the order parameter and
the diffusive dynamics. Next, we took on the challenge of studying the
finite momentum dynamics of the broken phase at zero magnetic field.
Because the simulation is at finite volume, the chiral condensate is not constant but wanders through the coset manifold at a finite rate.
We show that the diffusion coefficient for this process is consistent with dynamical scaling and is determined by the transport coefficients of a pion hydrodynamic EFT, which we review in \Sect{sec:broken}.  
Despite this subtlety, we were able to clearly observe the critical pion dynamics. The
culmination of this part is presented in Fig.~\ref{fig:phi_diffusion},
which compares the pion dispersion relation measured in this work to
the scaling predictions for the pion damping rate and velocity close to the critical point. The
agreement is striking and further exemplifies the precision of our
numerics.


Future directions to be explored are diverse. One avenue is to refine
the treatment of finite volume effects in the broken phase by developing a finite
volume real time EFT for the dynamics of the chiral condensate in the spirit of \cite{Hasenfratz:1989pk} and \cite{Yao:2022fwm}. The first steps in this direction
are presented in \App{sec:FokkerPlanck}, which worked out a Fokker-Planck description 
for the dissipative and Hamiltonian dynamics of the condensate  on the coset manifold. This provides a real time description of QCD in the $\epsilon$-regime,
suggesting connections with Random Matrix Theory~\cite{Damgaard:2011gc}.

A more physically motivated direction is the study of the Kibble-Zurek dynamics of this model. Physical systems are
typically not tuned at their physical point, but traverse the phase
transition, with their relevant coupling varying at a finite rate.
This finite rate competes with the critical relaxation rate of the
system and prevents the correlation length from  diverging.  When
transiting from the restored to the broken phase, this competition results in the
formation of finite-length domains with different values of the
condensate. A careful examination of this phenomenon, both in the
presence and absence of explicit symmetry breaking, will be crucial to
any future phenomenological predictions. We plan to utilize  the
precision measurements presented here and in our prior work~\cite{Florio:2021jlx}  to study this phenomenon in detail.

\subsection*{Acknowledgements}
The authors are grateful to J.~Pawlowski, F.~Rennecke, S.~Schlichting and L.~von~Smekal for interesting discussions and A. Soloviev and J. Bhambure for their collaboration on this work. This work was supported by the U.S. Department of Energy, Office of Science, Office of Nuclear Physics, Grants Nos. DE-SC0012704 (AF) and DE-FG88ER41450 (DT).

\appendix

\section{Hydrodynamics} 

This appendix extends the discussion of the hydrodynamics of the broken phase given in \Sect{sec:broken}. We will keep the temperature explicit in this section, following the practice of~\ref{sec:broken}.

\subsection{Correlation functions from the pion EFT}
\label{app:mean_field}

Here we will review the temporal correlation functions which arise
the from pion hydro equations of motion, \eq{eq:hydroeqns}, but include the
explicit symmetry breaking term and noise. Although these correlators are limits
of the mean field results derived in \cite{Grossi:2021gqi}, they
transcend mean field and follow directly from the hydrodynamic
equations of motion.  

The hydrodynamic equations (see \Sect{sec:broken}) for the phase and axial charge read
\begin{subequations}
\label{eq:hydroeqns2}
\begin{align}
\partial_t \vec{\varphi}  - \frac{\vec{n}_A}{\chi} =& \frac{\Gamma }{f^2} \, \left[ \nabla \cdot (f^2  \nabla \vec{\varphi}) - f^2 m^2 \vec{\varphi}) \right] + \vec{\xi}_\varphi \, ,  \label{eq:varphieqns2} \\
\partial_t  \vec{n}_A -  \nabla (f^2 \nabla \vec{\varphi}) + f^2 m^2 \vec{\varphi} =& D \nabla^2 \vec{n}_A + \vec{\xi}_A  \, ,
\end{align}
which follows from  the Poisson bracket between the phase and the charge, $\{\varphi_{s},n_{As'}\}=\delta_{ss'}$.
The vector charge remains diffusive
\st
\partial_t  \vec{n}_V  = D \nabla^2 \vec{n}_V  + \vec{\xi}_V \, .
\stp
\end{subequations}
The noise added to the right-hand-side of Eq.~\eqref{eq:hydroeqns2}  has variances given by the Fluctuation-Dissipation Theorem (FDT)
   \begin{subequations}
      \label{noise:all}
\begin{align}
   \llangle \xi_{\varphi, s}(t,\x) \xi_{{\varphi},s'}(t',\x') \rrangle =&  \frac{2 T\Gamma}{f^2} \delta_{ss'} \delta(t -t') \delta(\x -\x') \, ,  \\
   \llangle \xi_{As}(t,\x) \xi_{As'}(t',\x') \rrangle =& 2 T \sigma  \delta_{ss'} \delta(t -t') \left[-\nabla^2 \delta(\x -\x') \right] \, ,  \\
   \llangle \xi_{Vs}(t,\x) \xi_{Vs'}(t',\x') \rrangle =& 2 T \sigma  \delta_{ss'} \delta(t -t') \left[-\nabla^2  \delta(\x -\x') \right] \, .
   \label{noise:v}
   \end{align}
   \end{subequations}
where $\sigma=\chi_0 D$.

The equations are easily solved. For instance, we can Fourier transform in space and write the solution to the vector equation by integrating from $t=0$ up to $t$ (with $t> 0$)
\st
\vec{n}_V(t,\k) = \vec{n}(0,\k) e^{-Dk^2 t} + \int^t_{0} {\rm d}t_1\,   \vec{\xi}_V(t_1,\k) \,  e^{Dk^2 (t_1 - t)} \, .
\stp
The variance in the initial condition $\vec{n}(0,\k)$ is determined from the free energy functional in \eqref{eq:varphifree} and reads
\begin{align}
   \label{eq:diffusion_ic}
   \frac{1}{3V} \llangle \vec{n}_V(0, \k) \cdot \vec{n}_V(0,-\k) \rrangle =& T\chi_0 \, . 
\end{align}
Alternatively one can integrate from past infinity to determine the initial condition:
\st
     \vec{n}(0, \k) = \int^0_{-\infty} {\rm d}t_1\,   \vec{\xi}_V(t_1,\k) \,  e^{Dk^2 t_1 } \, .
\stp
Computing the variance of $\vec{n}(0,\k)$ using the variance of the noise in \eqref{noise:v}  reproduces \eqref{eq:diffusion_ic}, which demonstrates how the noise and dissipation work together in establishing the equilibrium state. By either method
\st
    \frac{1}{3V} \llangle \vec{n}(t, \k) \cdot \vec{n}(0,-\k) \rrangle  = T\chi_0 e^{-D k^2 t } \, .
\stp

An identical strategy can be used in the axial channel by finding the eigen functions of the system. As in the vector case, we Fourier transform in space and  calculate response function. It is convenient to work with  the vector of variables\footnote{For simplicity, $Y$ will denote a single isospin component of these fields.}  $Y = (y_1, y_2) = (\omega_k \varphi ,  \mu_{A})$ where $\mu_{A} =n_A/\chi$ and $\omega_k^2 = v^2(k +m^2)$.
The static susceptibility is diagonal in these variables,  and can be read off from the free energy written in \eqref{eq:varphifree}
\st
  \frac{1}{V} \llangle y_a(0,\k) \, y_b(0,-\k) \rrangle  = \frac{T}{\chi_0} \delta_{ab} \, .
\stp
As in the diffusion example, the static susceptibility sets the initial conditions for the subsequent evolution.
The equations of motion take the form $\partial_t Y + \mathcal M Y = \xi$ and
eigen-values of $\mathcal M$ are
\st
     \lambda_{\pm} = \pm i \omega_k + \frac{1}{2} \Gamma_k\, ,  \qquad \omega_k^2 \equiv v^2 (k^2 + m^2) \, , \qquad \Gamma_k \equiv \Gamma (k^2 + m^2)  + D k^2 \, .
\stp
The homogeneous solutions are $Y \propto e^{-\lambda_{\pm} t}$.
The full matrix of correlation functions
\begin{equation}
 [G_{\rm sym}(t,\k)] =
 \frac{1}{3V} \begin{pmatrix}
\omega_{k}^2  \langle \varphi_s(t,\k) \varphi_s(0,-\k) \rangle & \omega_{k}\langle \varphi_s(t,\k)\mu_{As}(0,-\k)  \rangle \\
 \omega_{k}\langle\mu_{As}(t,\k) \varphi_s(0,-\k)  \rangle & \langle  \mu_{As}(t,\k) \mu_{As}(0,-\k)\rangle
 \end{pmatrix} \, ,
\end{equation}
reads
\begin{equation}
 [G_{\rm sym}(t,\k)] = \frac{T}{\chi_0} e^{-\frac{1}{2}\Gamma_k t}
 \begin{pmatrix}
 \cos(\omega_k t )+ \frac{\Delta_k}{\omega_k}\sin(\omega_k t )& \sin(\omega_k t ) \\
-\sin(\omega_k t )&\cos(\omega_k t)- \frac{\Delta_k}{\omega_k}\sin(\omega_k t )
 \end{pmatrix}\ ,
\end{equation}
with $\Delta_k\equiv (Dk^2- \Gamma (k^2+m^2))/2$. In the body of the text this result is used in the massless limit in \eqref{eq:Gvphivphidynamic} and \eqref{eq:GAAeqn}.

\subsection{Diffusion of the chiral condensate on the coset manifold}
\label{sec:FokkerPlanck}

\subsubsection{ Free diffusion on the three sphere }
In a finite volume simulation the chiral condensate is not fixed
but diffuses on the coset manifold.
Let us examine this diffusion  by analyzing the zero mode of the hydrodynamic equations of motion given in \eqref{eq:hydroeqns2}, but limiting the discussion to $H=0$.  In finite volume the fields take the form
\begin{align}
   \vec{\varphi}(t, \x) =& V \vec{\varphi}_0(t) + \sum_{\k\neq 0} e^{i \k \cdot \x} \vec{\varphi}(t,\k) \, , \\
   \vec{n}_A(t, \x) =& \sum_{\k\neq 0} e^{i \k \cdot \x } \vec{n}(t,\k) \, .
\end{align}
The axial charge is exactly conserved and there is no net charge in the box, and consequently the $\k=0$ mode is zero for $\vec{n}_A$.
Then, integrating \eq{eq:varphieqns} over volume,  the zero mode of $\varphi$ satisfies a random walk
\st
  \partial_t \vec{\varphi}_0 =  \vec{\xi}_0 \, ,
 \qquad  \vec{\xi}_0(t)  \equiv  \int_\x \vec{\xi}_{\varphi}(t,\x)/V \, .
 \stp
Solving for $\varphi_0$  and computing its variance leads to \eq{eq:vevvariance}, which is  reproduced here for convenience
\st
  2 \, \mathfrak{D}\,  t \equiv  \frac{1}{3} \llangle \vec{\varphi}_0(t) \cdot \vec\varphi_0(t)  \rrangle = 2 \left(\frac{T\Gamma }{f^2 V}\right) t \, ,
\stp
i.e.  $\mathfrak{D} \equiv T\Gamma/f^2V$. 
The noise satisfies 
\st
\label{eq:variance}
  \llangle \xi_{\varphi 0s}(t) \xi_{\varphi 0s'}(t') \rrangle  = 2 \mathfrak{D}  \delta_{ss'} \delta(t -t') \, .
\stp
The  probability density $\mathcal P(\vec{\varphi}_0)$ satisfies the diffusion equation
\st
  \partial_t \mathcal P = \mathfrak{D} \frac{\partial}{\partial \varphi_{0s}} \left( \delta^{ss'} \frac{\partial \mathcal P}{\partial \varphi_{0s'} } \right) \, .
\stp

The direction of the chiral condensate is labeled by a unit vector on the three sphere $n_a n_a=1$.
We have parametrized this direction by  three angles,  $\varphi_{0s}$, assuming they are small.  The coordinates $\vec{\varphi}$ provide a local Cartesian coordinate patch for the three sphere close to the pole. In a general coordinate system the diffusion equation reads
\st
   \partial_t \mathcal P  = \mathfrak{D} \nabla_I \nabla^I \mathcal P  \, ,
\stp
where $\nabla_I\nabla^I$ is the Laplacian on the sphere.
More explicitly, a set of  coordinates covering the three sphere is 
$q^I = (q_1 , q_2, q_3) = (\varphi_{S}, \theta, \Phi)$ 
\st
\label{eq:threespherecoords}
  ds^2 = \mathrm{d}\varphi_S^2 + \sin^2\varphi_S \, (\mathrm{d}\theta^2 + \sin^2 \theta \, \mathrm{d}\Phi^2) \, ,
\stp
where $\varphi_S$ is approximately $\varphi_S \simeq \sqrt{\vec{\varphi}^2}$ close to the north pole. 

\subsubsection{Explicit symmetry breaking, forced diffusion,  and the $\epsilon$-regime of QCD}

When the symmetry is explicitly broken by a magnetic field,  the axial charge must be reconsidered. In this case the zero modes satisfy the equations of motion
\begin{align}
\partial_t \vec{\varphi}_0 -   \frac{\vec{n}_{A0}  }{\chi} =&  - \Gamma m^2 \vec{\varphi}_0 + \vec{\xi}_{\varphi_0} \, , \\
\partial_t \vec{n}_{A0} +   f^2 m^2 \vec{\varphi}_0 =&   0 \, .
\end{align}
where the statistics of $\vec{\xi}_{\varphi_0}$ are given in \eqref{eq:variance}.
The variables $\vec{\varphi}_0$ and $\vec{n}_{A0}$ are canonical conjugates $\{ \varphi_{0s}, n_{A0s'}\} = \delta_{ss'}$.  To illustrate the structure of these equations we introduce a traditional  notation
\st
( Q^s,  P_s) \equiv (\varphi_{0s}, n_{A0s} ) \, ,
\stp
and the equations of motion are written
\begin{align}
\partial_t Q^s + \{\mathcal H_0,  Q^s\} =& - \beta V \mathfrak{D}   \frac{\partial \mathcal H_0}{\partial Q^s} + \xi_0^s
\, ,\\
\partial_t P_s + \{\mathcal H_0,  P_s \} =&  0 \, .
\end{align}
Here $\beta$ and $V$ are the inverse temperature and volume respectively, and $\mathfrak{D}=T\Gamma/f^2V$ is the diffusion coefficient introduce earlier.
Finally the
 Hamiltonian  for the zero modes is
\st
 \mathcal H_0(Q,P) =  \frac{{\vec{P}}^{\,2}}{2\chi} +  \frac{1}{2} f^2 m^2 \vec{Q}^2  \, .
\stp

The Fokker-Planck equation for the phase space probability density $\mathcal P(Q, P)$
follows from  the Langevin dynamics and reads~\cite{kittel1961elementary,Arnold:1999va,Arnold:1999uza}
\st
 \partial_t \mathcal P + \left\{\mathcal P, \mathcal H_0 \right\} = \beta V\mathfrak{D} \,\frac{\partial}{\partial Q^s } \left( \delta^{ss'} \frac{\partial \mathcal H_0}{\partial Q^{s'}}\, \mathcal P\right) +
  \mathfrak D
 \frac{\partial}{\partial Q^s } \left( \delta^{ss'} \frac{\partial \mathcal P}{\partial Q^{s'}} \right) \, .
\stp
We emphasize that the  canonical measure  for the probability density $\mathcal P(Q,P)$ is invariant under canonical change of coordinates.

We have worked with a set of spatial coordinates,  which parametrize a region close to the pole of the three sphere by Cartesian coordinates. We then make a canonical change of variables to the polar coordinates $q^I$ given in \eqref{eq:threespherecoords}.
The Hamiltonian is generalized to include larger angles and reads
\st
   {\mathcal H}_0(q, p) =   \frac{g^{IJ}(q)}{2\chi} \, p_I p_J -  f^2 m^2 \cos(\varphi_S) \, ,
\stp
where $g^{IJ}(q)$ is the metric of the sphere  in \eqref{eq:threespherecoords}. 
Here we used the original free energy in \eqref{eq:Hdef} and the Gell-Mann, Oakes, Renner relation to deduce the potential for the zero mode 
\st
   \frac{-1}{V}\int_\x  \,  H_a \phi_a(t,\x)  = -H\bar \sigma  \cos\varphi_S=-f^2m^2 \cos(\varphi_S) \, .
\stp
The change of coordinates leads to the Fokker-Planck equation in its final form
\st
 \partial_t \mathcal P + \left\{\mathcal P, \mathcal H_0 \right\} =  \beta V \mathfrak D \,  \nabla_I \left(   \nabla^I\mathcal H_0 \,  \mathcal P \right) +
   \mathfrak D \, \nabla_I \nabla^I \mathcal P \, ,
\stp
where $\nabla_I$ is the covariant derivative on the three sphere.

It is straightforward to see that the steady state solution to this equation is the equilibrium probability distribution
\st
   \mathcal P(q,p)\,  {\mathrm d}^3 q \, \mathrm d^3 p = e^{-\beta V\mathcal H_0 } \mathrm{d}^3q \, \mathrm{d}^3p\, .
\stp
If the momenta are of no interest they can be integrated over yielding
\st
   \mathcal P(q)\,  \sqrt{g} \, {\mathrm d}^3 q \,  = e^{\beta Vf^2 m^2\cos\varphi_S } \sqrt{g}\, \mathrm{d}^3q \, ,
   \stp
which is consistent with the partition functions discussed in \cite{Hasenfratz:1989pk}.

It is satisfying that the dissipative dynamics of the finite volume chiral condensate on the three sphere is controlled by the (infinite volume) pion pole mass and damping coefficient.  
Indeed,  the Fokker-Planck equation provides a detailed real time picture of the so called ``epsilon" regime of QCD.  In this regime the quark mass is very small,  but 
$\beta Vf^2m^2 \sim 1$ is of order unity. The thermodynamics of this regime has been extensively studied using Random Matrix Theory and chiral perturbation theory (see \cite{Damgaard:2011gc} for an overview). It would be interesting to pursue this connection further.

\section{Static measurements in the $O(4)$ model}
\label{Sec:StaticMeasurements}

In this appendix we determine the correlation length and the magnetic susceptibility in the restored phase and along the critical line. In particular, their non-universal critical amplitudes serve to define universal ratios, including the novel dynamical ones introduced in the main text. We also determine the pion constant $f^2$ using static measurements.  For  convenience we collect the critical exponents used in this work in Table~\ref{tab:num_values_fixed}.
\begin{table}
  \centering
  \begin{tabular}{|c|c|c|c|}
  \hline
  exponent & definition & value & Ref. \\
  \hline
  $\beta$ &  $\bar{\sigma} \propto (-t_r)^{\beta}$ & 0.380(2)  & \cite{Engels:2014bra}\\
  \hline
  $\nu$ & $\xi \propto t_r^{-\nu}$ & 0.7377(41)  & \cite{Engels:2014bra}\\
  \hline
  $\gamma$ & $\chi \propto t_r^{-\gamma}$ & 1.4531(104) & \cite{Engels:2014bra}\\
  \hline
  \hline
  $\delta$ & $ \bar\sigma \propto H^{1/\delta}$ & 4.824(9) & \cite{Engels:2014bra}\\
  $\nu_c$ & $\xi_{L,T} \propto H^{-\nu_c}$ & 0.4024(2)  & \cite{Engels:2014bra}\\
  $1-1/\delta$ & $\chi_{L,T} \propto H^{-1 +1/\delta}$ & 0.7927(4)  & \cite{Engels:2014bra}\\
  \hline \hline
  $\omega$ & $Y \propto \xi^{\theta(Y)}(1+ c(Y) \xi^{-\omega} + \dots)$ & 0.755(5) &\cite{Hasenbusch:2021rse} \\
  $\zeta$ & $ \tau \propto \xi^{\zeta} $ & $d/2$ & \cite{Rajagopal:1992qz} \\ \hline
\end{tabular}
\caption{Critical exponents of the $O(4)$ model used in this work.
The top of the table is for zero field, $H=0$ and  $t_r=(m_0^2-m_c^2)/m_c^2$ is the reduced temperature. The middle of the table is on the critical line, $t_r=0$ and $H\neq0$.
The bottom of the table shows
the correction-to-scaling exponent $\omega$,   which is also universal.  In the table we denote ``any scaling quantity" by $Y$,  and $\theta(Y)$ and $c(Y)$ are its associated critical exponent and non-universal subleading amplitude respectively. $\zeta$ is the dynamical scaling exponent of ``Model G" where $d=3$ is the number of spatial dimensions.
}
\label{tab:num_values_fixed}
\end{table}

\subsection{Restored phase}
\label{app:corrlength_unbroken}

The magnetic susceptibility in the restored phase is defined as
\begin{equation}
  \chi(\k)= \frac{1}{4V}\sum_{a=0}^3 \left\langle|\phi_a(\k)|^2 \right\rangle \ ,
\end{equation}
and the zero momentum limit is notated with $\chi$
\begin{equation}
  \chi \equiv \chi(\bm 0) \ .
\end{equation}
A standard estimator of the correlation length is the so-called  ``second-moment correlation length'' \cite{Hasenbusch:2021rse}, computed as
\begin{equation}
\label{eq:second_moment}
  \xi = \sqrt{ \frac{\chi/F-1}{4\sin^2{\left(\frac{\pi}{L}\right)}} } \ ,
\end{equation}
with $F$ the susceptibility of the first Fourier mode
\begin{equation}
  F=\chi\left(\k\right)|_{k= \frac{2\pi}{L}}.
\end{equation}

\begin{figure}
  \centering
  \subfloat{\includegraphics[scale=1]{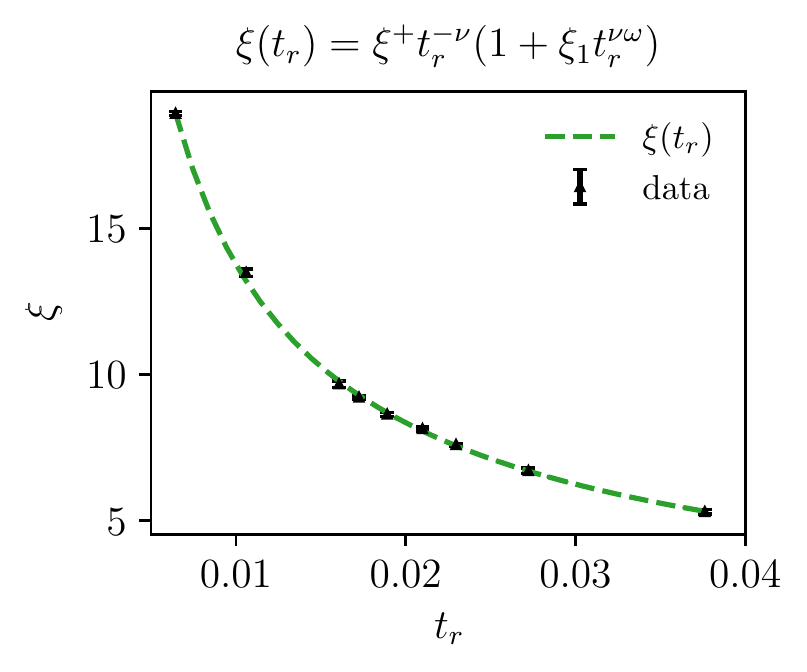}}
  \subfloat{\includegraphics[scale=1]{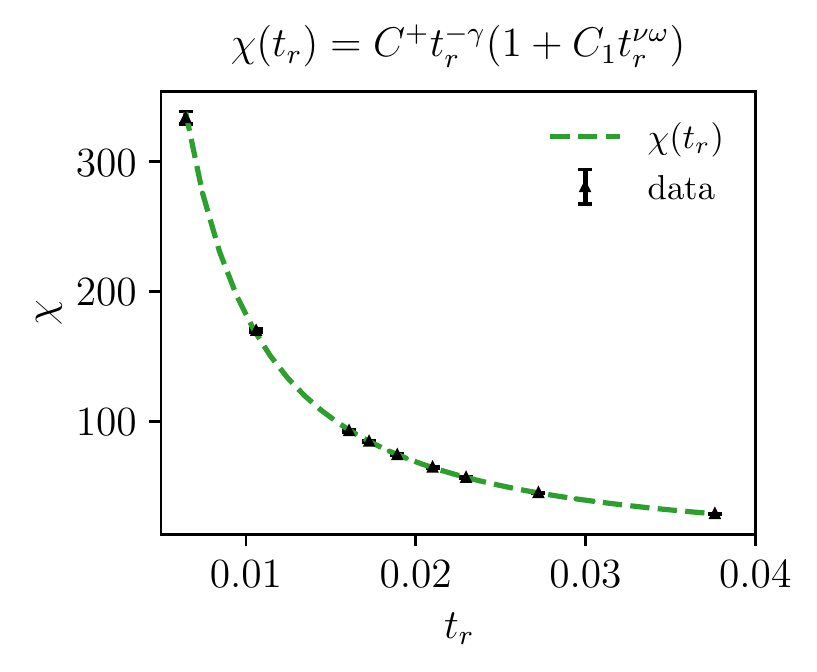}}
  \caption{\textbf{Left:} Correlation length as a function of the reduced temperature.
  The fitted non universal amplitude is  $\xi^{+}= 0.4492\pm0.0036$ and the subleading amplitude is $\xi_1= 0.33\pm0.09$.
  The $\chi^2/\mathrm{dof}$ is $9.1/7$.
   \textbf{Right:} Magnetic susceptibility as a function of the reduced temperature.
   The fitted non universal amplitude is $C^{+}= 0.208\pm0.003$  and the subleading correction
   is $C_{1}=1.11\pm0.15$.
    The $\chi^2/\mathrm{dof}$ is $4.83/7$.
    In both cases, the error bars are too small to be visible on the plot.}
  \label{fig:static_fit_restored}
\end{figure}

We fit their dependence on the reduced temperature $t_r=(m_0^2 -m^2_c)/{m_c^2}$ to
\begin{align}
  \xi(t_r) &= \xi^+t_r^{\nu}(1+\xi_1 t_r^{\omega\nu}) \ , \\
  \chi(t_r) &= C^+t_r^{\gamma}(1+C_1 t_r^{\omega\nu}) \ .
\end{align}
The exponent $\omega$ is the first subleading scaling exponent. To perform the fit, we fix $\beta$, $\nu$ and $\omega$ to the values presented in Tab.~\ref{tab:num_values_fixed} and fit $\xi^+$, $\xi_1$, $C^+$ and $C_1$. The fits are presented in Fig.~\ref{fig:static_fit_restored}. We obtain
\begin{equation}
\xi^+ = 0.4492\pm0.0036, \ \ \xi_1 = 0.332\pm0.085, \ \ C^+ = 0.2077\pm0.0032, \ \ C_1=1.11\pm0.15 \ .
\end{equation}
The chi-square for the fit of the correlation length is $\chi^2/\mathrm{dof}=9.1/7$ and the one for the susceptibility is $\chi^2/\mathrm{dof}=4.83/7$.
These values, together with the value the non-universal value for the magnetization in the restored phase
\begin{equation}
  \bar\sigma(t_r) = B^- (-t_r)^\beta(1+B_1 (-t_r)^{\nu\omega} ) \ ,
\end{equation}
that was extracted in Ref.~\cite{Florio:2021jlx}\footnote{Note an unfortunate misprint. Equation (29) in the published version should read $\Sigma=b_1(m_c^2-m_0^2)^\beta(1+C_T(m_c^2-m_0^2)^{\omega\nu})$. The then quoted parameters $b_1=0.544\pm0.004$ and $C_T =0.20\pm0.02$ lead to $B^-=|m_c^2|^\beta b_1 = 0.988\pm0.007$ and $B_1=C_T|m_c^2|^{\omega\nu} = 0.49\pm0.05$ .
}
\begin{equation}
  B^-=0.988 \pm 0.007 \, ,
\end{equation}
allow us to compute the universal ratio \cite{Engels:2003nq}
\begin{equation}
  Q_c = \frac{(B^-)^2\xi^3}{C^+}  = 0.4266 \pm 0.0126 \, .
\end{equation}
This value is indeed compatible with the $Q_c=0.4231(9)$ of \cite{Engels:2003nq}.

\subsection{Critical line $m_0^2=m_c^2$}
\label{app:corrlengthcrit}

\begin{figure}
  \centering
  \subfloat{\includegraphics{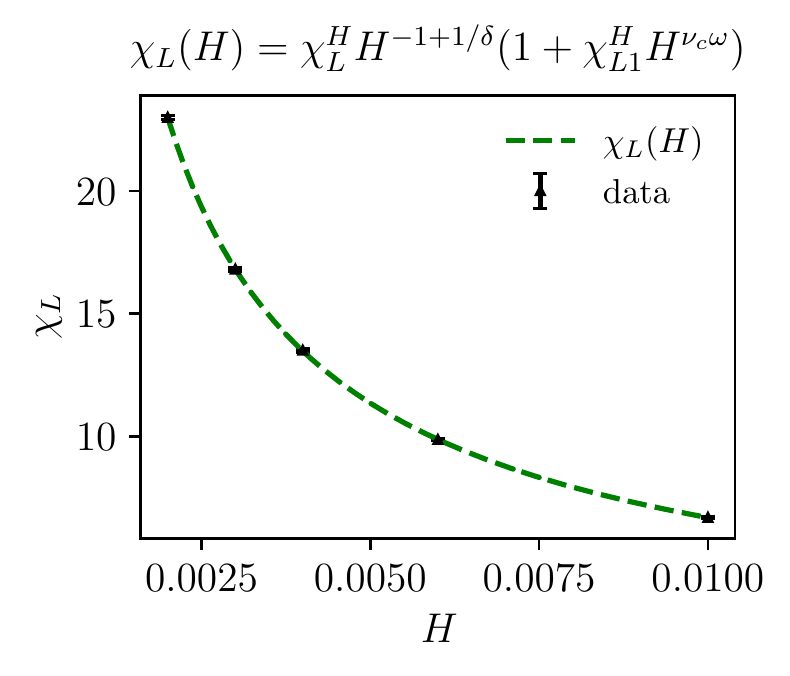}}
  \subfloat{\includegraphics{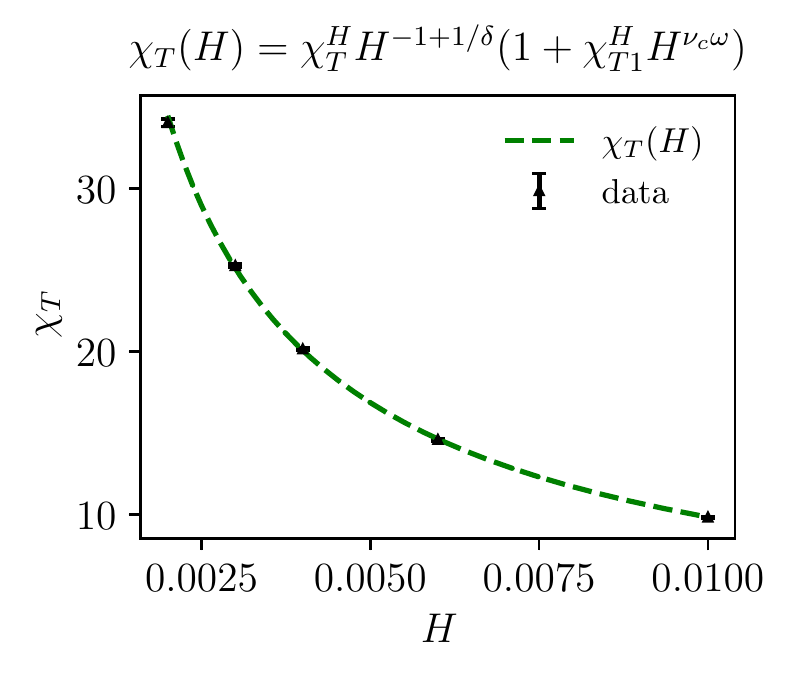}}
  \caption{\textbf{Left}:
 	Longitudinal susceptibility $\chi_L$ as a function of the magnetic field $H$.
	The data points are fit with $\chi_L(H)=\chi_{L}^HH^{-1+1/\delta}(1+\chi^H_{L1} H^{\nu_c \omega})$. The results are
$\chi_{L}^H =0.1564  \pm 0.0016$ and  $\chi^H_{L1}=0.45 \pm 0.06$ with $\chi^2/\mathrm{dof}=0.26/3$.
   \textbf{Right}: Perpendicular susceptibility $\chi_T$ as function of the magnetic field $H$. The data points are fit with $\chi_T(H)=\chi^H_{T}H^{-1+1/\delta}(1+\chi^H_{T1} H^{\nu_c \omega})$.
   The results are
$\chi_{T}^H =0.241 \pm 0.004$ and  $\chi^H_{T1}=
0.25 \pm 0.08$ with $\chi^2/\mathrm{dof}=6.17/3$.
}
  \label{fig:static_chi}
\end{figure}

At a non-zero magnetic field the magnetization will be parallel to the magnetic field. Then it is natural to define and measure the longitudinal and transverse $\chi_L$ susceptibilities:
\begin{align}
\chi_L(\k)&= \frac{1}{V} \langle |\phi_0(\k)|^2 \rangle_c \, ,
\\
\chi_T(\k) &=  \frac{1}{3V}\sum_{s=1}^3 \langle |\phi_s(\k)|^2 \rangle_c \, .
\end{align}
The static susceptibilities are shown in \Fig{fig:static_chi}. As a fitting function, we chose the scaling prediction with the first subleading correction
\begin{equation}
\chi(H)= \chi H^{-1+\frac{1}{\delta} } (1 + \chi_1 H^{\nu_c \omega} ) \, .
\end{equation}

\begin{figure}
  \centering
  \subfloat{\includegraphics{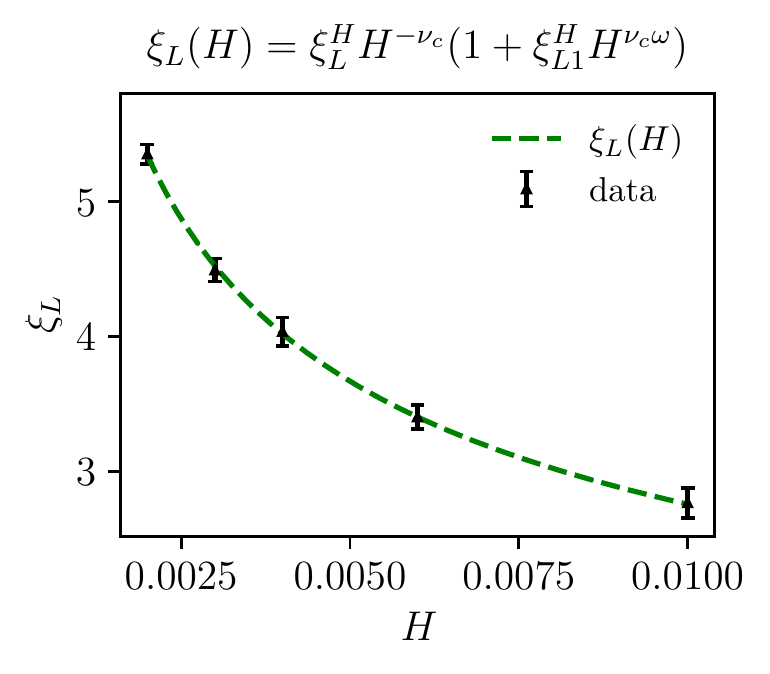}}
  \subfloat{\includegraphics{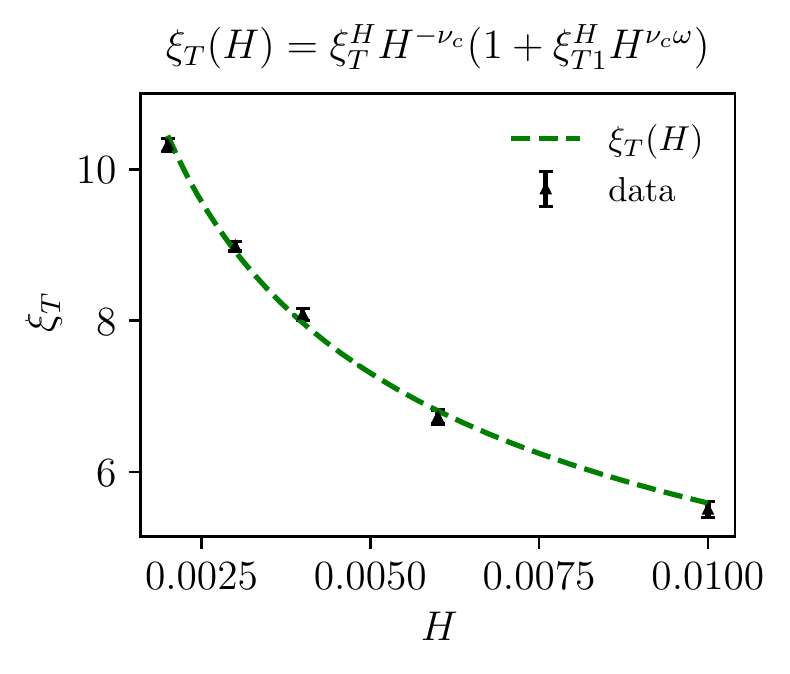}}
  \caption{\textbf{Left:} Longitudinal correlation length $\xi_L$ as function of the of the magnetic field $H$ .
  The data points are fit with $\xi_L=\xi_{L+}H^{-\nu_c}(1+\xi_{L1} H^{\nu_c \omega})$. The results are
$\xi_{L+} =0.447\pm0.026$ and  $\xi_{L1}=-0.14 \pm 0.32$ with $\chi^2/\mathrm{dof}=0.19/3$.
   \textbf{Right}: Perpendicular correlation length $\xi_T$ as function of the magnetic field $H$. The data points are fit with $\xi_T=\xi_{T+}H^{-\nu_c}(1+\xi_{T1} H^{\nu_c \omega})$.
   The results are
$\xi_{T+} = 0.826 \pm 0.027$ and  $\xi_{T1}=0.24 \pm 0.19$ with $\chi^2/\mathrm{dof}=6.76/3$.}
  \label{fig:static_xi}
\end{figure}

To extract the correlation length in the longitudinal and the transverse directions we measure the second-moment correlation length
as in \eqref{eq:second_moment} with
 \begin{align}
F_L&= \chi_L\left(\k\right)|_{k= \frac{2\pi}{L}}
 \, , \\
F_T &=\chi_T   \left(\k\right)|_{k= \frac{2\pi}{L}} \, .
\end{align}
The scaling prediction fixes the fitting function as
\begin{align}
\xi_L(H)&=\xi_{L}^HH^{-\nu_c}(1+\xi_{L1}^H H^{\nu_c \omega}) \, , \\
\xi_T(H)&=\xi_{T}^HH^{-\nu_c}(1+\xi_{T1}^H H^{\nu_c \omega}) \, ,
\end{align}
and the result of the fit is shown in Fig.~\ref{fig:static_xi}.
The universal ratio $\xi_{T+}/\xi_{L+}=1.85\pm0.12$ is consistent with the more precise estimate done in \cite{Engels:2003nq}.

\subsection{Broken phase and the pion velocity}
\label{app:staticanalysis}

Our goal in this appendix is to extract the infinite volume chiral condensate $\bar\sigma(T)$ and pion velocity $v^2(T) = f^2(T)/\chi_0$ by analyzing static correlators below the critical point at $H=0$. The resulting parameterization of the velocity is shown in \Fig{fig:phi_diffusion} ({Right}). We will use the magnetization
\st
M_a(t) \equiv \frac{1}{V} \sum_{\x} \phi_a(t,\x) \, ,
\stp
the static correlation function $G_{\phi\phi}(0,\k)$, and their
their dependence on volume to extract $\bar\sigma(T)$ and $(\bar \sigma^2/f^2)(T)$  in the infinite volume limit.

In infinite volume the  magnetization remains constant in time and determines
the chiral condensate
\st
  \lim_{V \rightarrow \infty} M_a = \bar\sigma(T) \, n_a  \, .
\stp
Here $n_a$ is an arbitrary unit vector on the three-sphere characterizing the orientation of the condensate.
However, in any finite volume
\st
    \llangle M_a \rrangle=0 \, ,
\stp
since the condensate orientation vector $n_{a}$ stochastically explores the $S_3$ in time -- see \App{sec:FokkerPlanck}.

One way to extract  $\bar\sigma$ is to look
at the fluctuations of $M_a$,  evaluating $\llangle M^2 \rrangle = \llangle M_a M_a \rrangle$,
  which is approximately $\bar\sigma^2$ at large volumes.
The leading deviation of $\llangle M^2 \rrangle$ from $\bar\sigma^2$ at finite volume comes from the fluctuations of long wavelength Goldstone modes and can be neatly analyzed with a Euclidean pion effective theory~\cite{Hasenfratz:1989pk,Dimitrovic:1990yd}
In the chiral limit the only parameter of the EFT is the pion  constant  $f^2(T)$. When a finite magnetic field is included
the chiral condensate $\bar\sigma$ is also a parameter.

Apart from a slightly different fitting procedure, the current determination of ${\bar\sigma (T)}$ simply repeats the analysis done in our earlier work for the larger lattices and datasets used here~\cite{Florio:2021jlx}.
The expansion relating $\llangle M^2 \rrangle$ and $\bar \sigma^2$  takes the form~\cite{Hasenfratz:1989pk,Florio:2021jlx}
\st
 \label{eq:hasenfratznumeric1}
 \llangle M^2\rrangle = \bar\sigma^2 \left[  1   +\frac{0.677355}{f^2 L} + \frac{0.156028}{f^4 L^2}  + O\left((f^2L)^{-3}\right) \right] \, ,
\stp
and is valid for $f^2(T) L\gg 1$. The numerical coefficients are known analytically in terms of ``shape coefficients" $\beta_1$ and $\beta_2$, but are of no interest here.

To determine the chiral condensate, we plotted $\llangle M^2 \rrangle$ versus $L$ at fixed temperature, and made various fits with \eqref{eq:hasenfratznumeric1}
to determine $\bar\sigma^2(T)$  and $(\bar\sigma^2/f^2) (T)$.
We estimated the systematic uncertainty in the extracted values of  $\bar\sigma(T)$ by adding a $C/L^3$ term to the fit form and comparing to a fit based on \eqref{eq:hasenfratznumeric1}.
Except for the point closest to $T_c$ the systematic uncertainty $\bar\sigma(T)$ is smaller than
our statistical uncertainty, which would not have been the case if only  the first  term $\propto 1/f^2L$  in the expansion  had been used.

\begin{figure}
\centering
\includegraphics[scale=0.99]{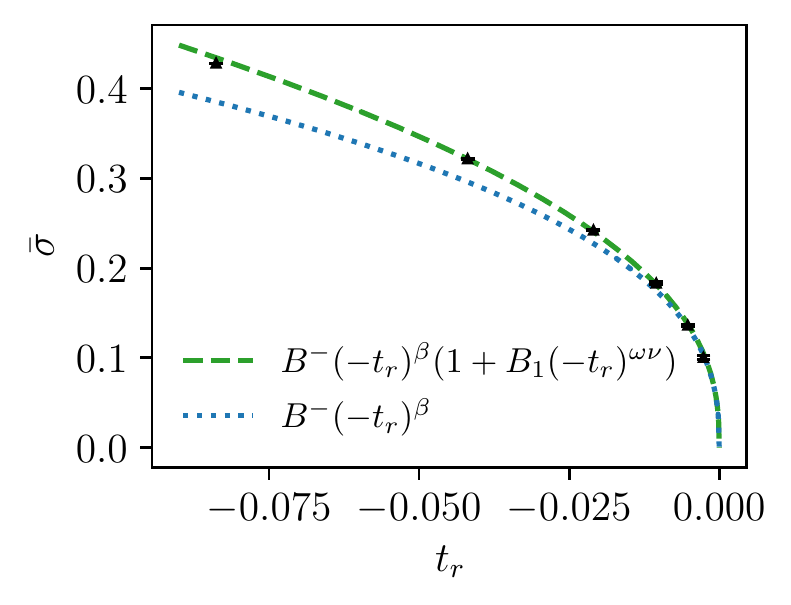}
\includegraphics[scale=0.99]{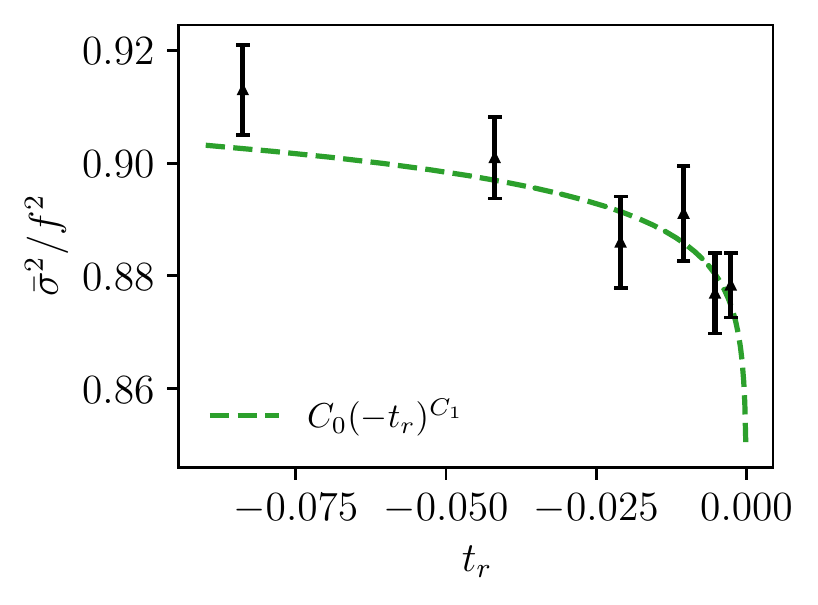}
\caption{\textbf{Left:} Magnetization in the broken phase for zero magnetic field after an extrapolation to infinite volume developed in \cite{Hasenfratz:1989pk}. The dashed line is  a fit to the data   with
$ \bar\sigma(t_r)= B^{-} (-t_r)^\beta(1 + B_1 (-t_r)^{\omega \nu})$ including  the subleading correction. The dotted line shows the leading term of the  fit.
\textbf{Right:}
 An  estimate of the pion constant $f^2$ (or more precisely $\bar\sigma^2/f^2$) based on static correlations, together with a phenomenological fit ansatz.}
\label{fig:derekbrokenshenanigans}
   \end{figure}

Our results for $\bar\sigma(T)$  are shown in the right panel of  Fig.~\ref{fig:derekbrokenshenanigans},
This is fit with the functional form
\st
\label{eq:sigmavst}
\bar \sigma(t_r) =  B^{-} (-t_r)^{\beta} \left(1 +  B_1 (-t_r)^{\omega \nu}  \right) \, .
\stp
with fit parameters $B^{-} =0.988\pm 0.007$, $B_1=0.51 \pm 0.07$.
The results for the vev amplitude and subleading corrections are compatible  with our previous results.
For comparison,
we also show the fit results for the first term $B_1(-t_r)^\beta$. Clearly,
for precision work, the subleading corrections are important in the temperature
range we are considering.


The decay constant $f^2$ of the pion EFT can be extracted using similar methods
The correlation function of the scalar field
at small wavenumbers
\st
  k \sim \frac{1}{L}  \ll m_\sigma \ll \frac{1}{a} \, ,
\stp
is dominated by the massless Goldstone mode.
Recalling \Eq{eq:Gphiphistatic} with $T=1$,
the correlation function takes the form
\st
   \frac{4}{3} G_{\phi \phi}(0,{\bm k}) =   \frac{\bar\sigma^2}{f^2 k^2} \ ,
\stp
to leading order in the pion EFT.
However,  there are  important corrections to this result stemming from the finite system size, which  are of order $1/f^2 L$ and  are captured by the leading order pion EFT. Short distance correlations of order $(m_{\sigma} L)^2$ and smaller  lead to higher derivative terms in the pion EFT, which are not included. Naturally, these terms become increasingly important near the critical point.
Following the perturbative expansion of  Hasenfratz and Leutwyler the correction takes the form
  \st
   \frac{4}{3} G_{\phi\phi}(0,{\bm k})  =  \frac{\bar \sigma^2}{f^2k^2}  \left[1 + \frac{\beta_1}{f^2 L} -\frac{N-2}{f^2 k^2 V}   + \mathcal O\left(\frac{1}{(m_\sigma^2 L)^2}\right) \right] \, .
  \stp
  where $\beta_1 = 0.225785$ is a ``shape coefficient" reflecting the fluctuations of pions in a cubic box.
Numerically for $k=2\pi/L$ and $N=4$:
\begin{align}
   \label{eq:hasenfratznumeric2}
    \frac{4}{3} k^2 G_{\phi\phi}(0,{\bm k}) \Big|_{k=2\pi/L} =&  \frac{\bar\sigma^2}{f^2} \left(1 + \frac{0.175124}{f^2 L} \right) \, .
\end{align}
To extract $\bar\sigma^2/f^2$ we plotted $k^2 G_{\phi\phi}(0,{\bm k})$ for the first Fourier mode as a function of system size and fitted the result for $(\bar\sigma^2/f^2)(T)$ and $1/f^2$ using \eq{eq:hasenfratznumeric2}. The subleading correction $\sim 1/f^2 L$ in \eqref{eq:hasenfratznumeric2} was treated as a free parameter, although it is broadly consistent with the $\bar\sigma^2(T)$ extracted above and the $(\bar\sigma^2/f^2)(T)$ determined here.

The data on $(\bar\sigma^2/f^2)(T)$ are shown in Fig.~\ref{fig:derekbrokenshenanigans}. Asymptotically, $\bar\sigma^2/f^2$ should behave
as
\st
  \frac{\bar\sigma^2}{f^2} \sim   (-t_{\rm r})^{\eta} \, ,
\stp
where $\eta = 0.030$ is a critical exponent. We are unable to extract $\eta$ from our measurements. Instead, we have simply fit a form
\st
\label{eq:fitresultsS2overF2}
  \frac{\bar\sigma^2}{f^2} = C_0 (-t_{\rm r})^{C_1} \, ,
\stp
with $C_0=0.910(8)$ and $C_1=0.009(3)$, which gives a heuristic parametrization of the results shown in the figure. There is  a hint in the data that $\eta$ is non-zero.

Given the fit results for $\bar\sigma^2(T)$ in \eqref{eq:sigmavst} and \eq{eq:fitresultsS2overF2} for $(\bar\sigma^2/f^2)(T)$ we can extract the velocity
\st
     v^2 = \frac{f^2}{\chi_0} = \frac{1}{\chi_0} \frac{\bar\sigma^2(T)}{(\bar\sigma^2/f^2)(T) } \, .
\stp
The uncertainty is dominated by the contribution from $\bar\sigma^2(T)$ as $\bar\sigma^2/f^2$ (which is nearly unity) in total provides a ten percent correction. In Fig.~\ref{fig:derekbrokenshenanigans} we have varied the fit parameters in the range allowed by the fits to deduce the static velocity and its uncertainty.

\section{Axial channel systematics in the broken phase}
\label{app:axial_channel}

We record in this appendix our systematic analysis leading to the extraction of $\Gamma(k)$ and $\omega(k)$ in the broken phase.

\begin{figure}
  \centering
  \subfloat{\includegraphics{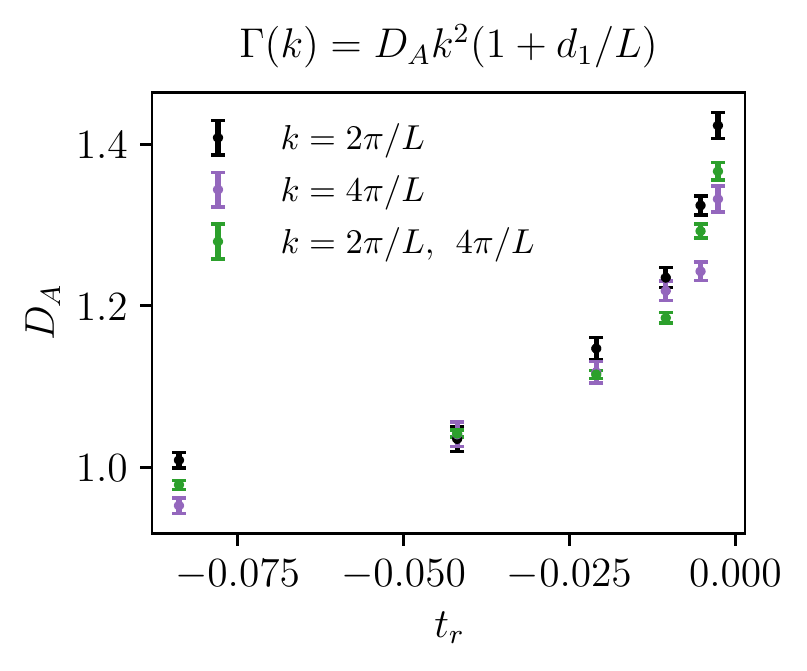}}
  \subfloat{\includegraphics{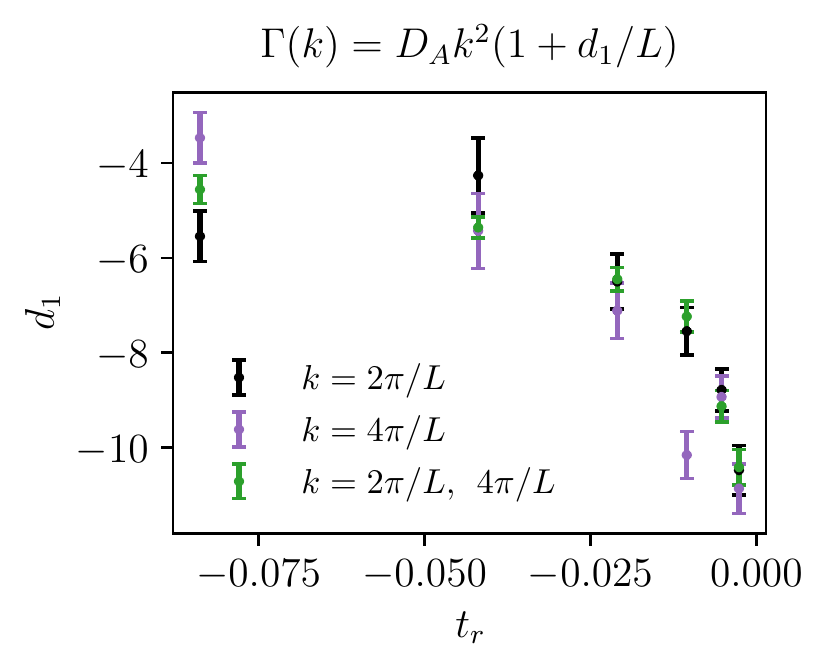}}
  \caption{Result of the fit of the damping rate $\Gamma(k)=D_Ak^2(1+ \frac{d_1}{L})  $.
  \textbf{Left:}
$D_A$ coefficient as a function of $t_r$ extracted for $k=2\pi/L$ (black dots), $k=4\pi/L$ (purple dots) and the two datasets together (green dots).
   \textbf{Right:}
   $d_1$ coefficient as a function of $t_r$ extracted for $k=2\pi/L$ (black dots), $k=4\pi/L$ (purple dots) and the two datasets together (green dots). }
  \label{fig:da_fit}
\end{figure}

To validate the fitting form (\ref{eq:fittingform})
\begin{align}
\Gamma(k)&= D_A k^2(1+d_1 /L)\, ,\\
\omega(k)&=v k + v_2 k^2 \, ,
\end{align}
and  the specific dependence on the size $L$ of the axial diffusion coefficient $D_A$,  we fit the damping coefficient $\Gamma$ as function of the reduce temperature $t_r$
using three different datasets:  $k=2\pi/L$, $k=4\pi/L$ and the combined sample.
Figure~\ref{fig:da_fit}  shows the results obtained for $D_A$ and the leading volume correction $d_1$ across the three different datasets.
The consistency of the fit among the different datasets and the values of $d_1$ exemplify the importance of the size-dependent $d_1/L$ term in (\ref{eq:fittingform}) in the axial diffusion coefficient.
Conversely, the remaining systematic dispersion between the different datasets can be considered as  an evaluation of the remaining systematic errors.

\begin{figure}
  \centering
  \subfloat{\includegraphics{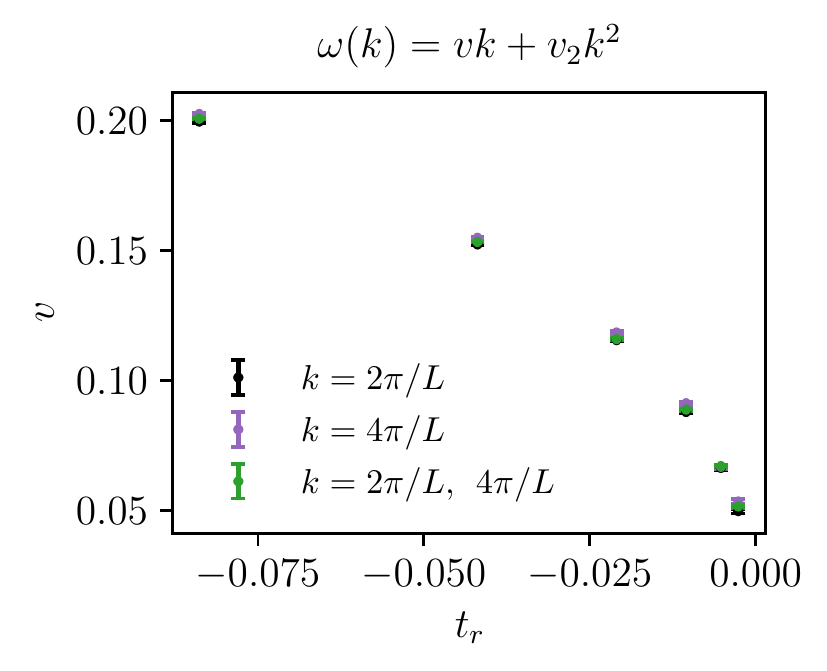}}
  \subfloat{\includegraphics{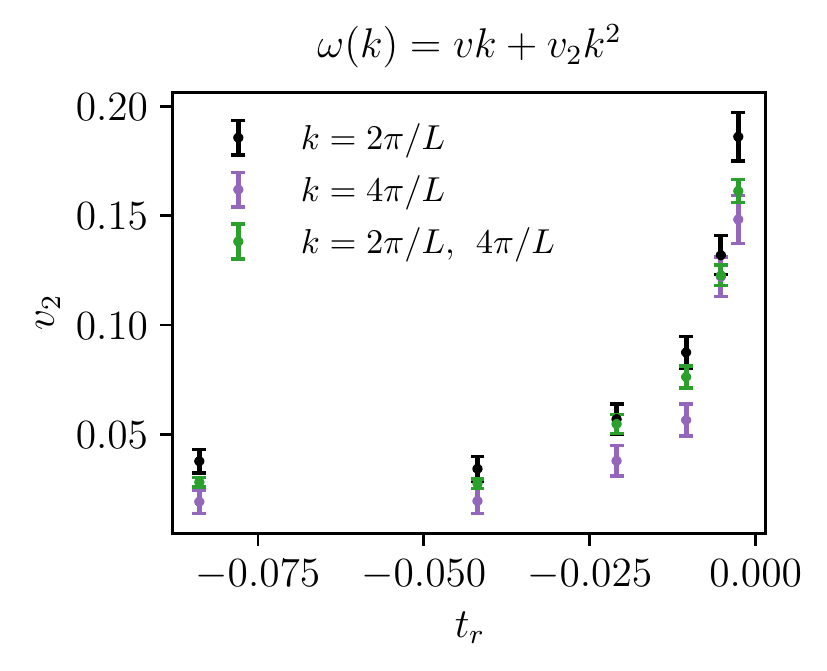}}
  \caption{
  Result of the fit of the dispersion curves $\omega(k)=vk+v_2k^2 $.
  \textbf{Left:}
$v$ coefficient as a function of $t_r$ extracted for $k=2\pi/L$ (black dots), $k=4\pi/L$ (purple dots) and the two datasets together (green dots).
   \textbf{Right:}
   $v_2$ coefficient as a function of $t_r$ extracted for $k=2\pi/L$ (black dots), $k=4\pi/L$ (purple dots) and the two datasets together (green dots).
   }
  \label{fig:omega_da}
\end{figure}

The same strategy is adopted for the extraction of the parameter for the dispersion curves, $\omega(k)$ that is fitted as
\begin{equation}
\omega(k)=v k + v_2 k^2 \ ,
\end{equation}
where the quadratic term $v_2 k^2$ was needed to improve the quality of the fit.
In Figure~\ref{fig:omega_da} we show the results for the coefficient $v$ and $v_2$.
We see that in this case, the systematic errors are smaller and that the fit agrees very well across all different datasets.

Let us conclude this appendix  by commenting on our assessment of statistical errors for this analysis. We subdivided each simulation in the time direction into blocks and fit each block independently. We then used  the mean and the standard error of this set of block fits as an estimator for the best fit parameters and their statistical error.

\clearpage

\bibliography{refs}

\end{document}